\documentclass[aps,prx,superscriptaddress,twocolumn]{revtex4-2}

\usepackage{amsmath,amssymb}
\usepackage{graphicx}
\usepackage{xcolor}
\usepackage[colorlinks=true,allcolors=blue]{hyperref}

\newcommand{\streda}{St\v{r}eda }
\newcommand{\fvec}[1]{\boldsymbol{#1}}
\newcommand{\half}{\frac{1}{2}}
\newcommand{\rmd}{{\rm d}}

\begin{document}

\title{Extended Fractional Chern Insulators Near Half Flux in Twisted Bilayer Graphene Above the Magic Angle}

\def\StanfordPhys{Department of Physics, Stanford University, Stanford, CA 94305}
\def\SIMES{Stanford Institute for Materials and Energy Sciences, SLAC National Accelerator Laboratory, Menlo Park, CA 94025}
\def\StanfordAP{Department of Applied Physics, Stanford University, Stanford, CA 94305}
\def\NHMFL{National High Magnetic Field Laboratory, Tallahassee, Florida, 32310, USA}
\def\NIMS{Research Center for Functional Materials, National Institute for Materials Science, 1-1 Namiki, Tsukuba 305-0044, Japan}
\def\ICMN{International Center for Materials Nanoarchitectonics, National Institute for Materials Science,  1-1 Namiki, Tsukuba 305-0044, Japan}
\def\MIT{Department of Physics, Massachusetts Institute of Technology, Cambridge, MA 02139}
\def\FSU{Department of Physics, Florida State University, Tallahassee, Florida 32306, USA}
\def\SHTEC{School of Physical Science and Technology, ShanghaiTech University, Shanghai, 200031, China}

\author{Joe~Finney}
\thanks{Equal contributor with $\dagger$.}
\affiliation{\StanfordPhys}
\affiliation{\SIMES}

\author{Aaron~Sharpe}
\email{Equal contributor with $*$; aaron.sharpe@stanford.edu}
\affiliation{\StanfordPhys}
\affiliation{\SIMES}

\author{Linsey~K.~Rodenbach}
\affiliation{\StanfordPhys}
\affiliation{\SIMES}

\author{Jian~Kang}
\affiliation{\SHTEC}

\author{Xiaoyu~Wang}
\affiliation{\NHMFL} 

\author{Kenji~Watanabe}
\affiliation{\NIMS}

\author{Takashi~Taniguchi}
\affiliation{\ICMN}

\author{Marc~A.~Kastner}
\affiliation{\StanfordPhys}
\affiliation{\SIMES}
\affiliation{\MIT}

\author{Oskar~Vafek}
\affiliation{\NHMFL}
\affiliation{\FSU}

\author{David~Goldhaber-Gordon}
\affiliation{\StanfordPhys}
\affiliation{\SIMES}

\date{\today}

\begin{abstract}
Fractional Chern insulators (FCIs)---the lattice analog of fractional quantum Hall states---form as fractionalized quasiparticles emerge in a partially-filled Chern band. This fractionalization is driven by the interplay of electronic interaction and quantum geometry of the underlying wavefunctions. 
Bilayer graphene with an interlayer twist near the magic angle of 1.1\textdegree\  hosts diverse correlated electronic states at zero magnetic field. 
When the twist angle exceeds 1.3\textdegree, the electronic bandwidth is sufficient to suppress the zero-field correlated states.
Yet applying a magnetic field can restore the importance of electron-electron interactions. Here, we report strongly-correlated phases when a 1.37\textdegree\ twisted bilayer graphene sample is tuned to near half a magnetic flux quantum per moir\'e cell, deep into the Hofstadter regime. 
Most notably, well-quantized odd-denominator FCI states appear in multiple Hofstadter subbands over unusually large ranges of density.
We also observe a bending and resetting of the Landau minifan reminiscent of behavior commonly seen in magic-angle samples near integer filling at low magnetic field. 
\end{abstract}

\maketitle

\section*{Introduction}

Twisted bilayer graphene (TBG), the archetypal strongly-correlated moir\'e heterostructure, displays astonishingly diverse correlated electronic and topological phases~\cite{Cao2018,Cao2018_2, Yankowitz2019,Sharpe2019,Andrei2020, Balents2020,Pierce2021,Xie2021,Saito2021,Liu2021,Lu2021,Stepanov2021}. The electronic moir\'e miniband structure depends sensitively and predictably on twist angle. Near the magic angle of 1.1\textdegree, the large moir\'e length scale enables tuning the density of charge carriers through entire minibands and the density of magnetic flux to on order one flux quantum $\Phi_0=e/h$ per moir\'e cell, where $e$ is the elementary charge and $h$ is Planck's constant. In this ``Hofstadter's butterfly'' regime~\cite{Hofstadter1976,Bistritzer2011_2,Dean2013,Hunt2013,Ponomarenko2013,Wang2015,Spanton2018,Lu2021,das_observation_2022,Wang2022,Yu2022,herzog-arbeitman_reentrant_2022,Wang2024}, the energy spectrum exhibits a fractal structure with field-induced topological subbands. Gaps in the fractal spectrum appear at fluxes and densities described by Diophantine equations of the form $n/n_s=s + t\Phi/\Phi_0$~\cite{Wannier1978}. Here, $n_s=1/A$, $\Phi=BA$, $A$ is the moir\'e unit cell area, $B$ the magnetic field normal to the plane of the samples, $s$ the density offset at zero flux, and $t$ the Chern number associated with the gap. We notate these \streda lines as $(s,\  t)$. Within such a gap, the Hall conductance is expected to be quantized to $\sigma_{xy} = te^2/h$~\cite{Thouless1982}.

Integer $t$ \streda lines have been observed in TBG at a range of twist angles. Some of these states are not describable by single-particle models, as symmetries are broken by electronic interactions~\cite{Young2012, Pierce2021, Stepanov2021, Yu2022, Wang2024}. 
\streda lines with fractional $t$ and nonzero $s$ are more exotic and are referred to as fractional Chern insulators (FCIs)~\cite{kol_fractional_1993, neupert_fractional_quantum_2011, neupert_fractional_2011, regnault_fractional_2011, santos_time-reversal_2011, sheng_fractional_2011, sun_nearly_2011, tang_high-temperature_2011}. Analogous to fractional quantum Hall (FQH) states, where $s=0$, FCIs host fractionalized excitations~\cite{Laughlin1983}.
Only a few examples of moir\'e materials hosting such states have been experimentally reported, in twisted MoTe$_2$ ~\cite{Cai2023, Zeng2023, Park2023} and multilayer twisted graphene aligned with hexagonal boron nitride (hBN) ~\cite{Spanton2018, Xie2021, Lu2024, luExtendedQuantumAnomalous2025}.

TBG's viability as a platform for FCIs remains an open question~\cite{Parker2021}. Recently, fractional $t$ states have been observed in high-quality magic-angle TBG~\cite{he_strongly_2024}.
These occurred in Landau levels originating from charge neutrality ($s=0$), reminiscent of ordinary FQH. So far, FCIs with $s\neq 0$ have not been clearly demonstrated in experiments on TBG devices not aligned to hBN.

\begin{figure*}[!t]
\centering
\includegraphics[width=7.2in]{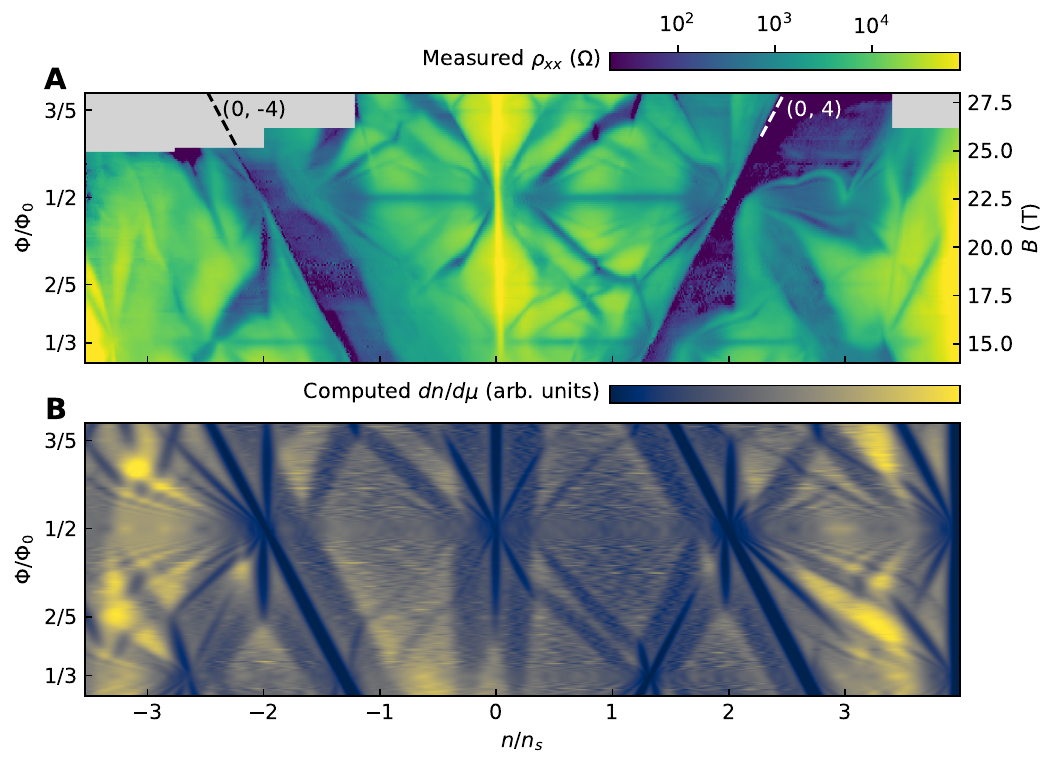}
\caption{\textbf{Strained TBG in the Hofstadter butterfly regime.} (\textbf{A}) 
Dependence of the longitudinal resistivity as a function of density and field. We indicate \streda lines $(s,\ t)=(0,\, \pm4)$ with dashed lines. The metallic top gate began to leak to the TBG above 25~T, so we reduced its voltage range, leaving the grayed-out regions. See the Supplement for a schematic description of quantum Hall gaps (Fig.~\ref{fig:megafan_sup}), the behavior of other contact pairs (Figs.~\ref{fig:all_xx} and \ref{fig:all_xy}), a comparison with the lower-field data presented in Ref.~\cite{Finney2022} (Fig.~\ref{fig:tlcomp}), and a schematic of the device with our inferred twist angles as extracted from the Landau fan diagram (Fig.~\ref{fig:schematic}). (\textbf{B}) Computed density of states d$n$/d$\mu$ for the spectrum in Fig.~\ref{fig:computation}A. Gaps show up as dark lines. See Sec.~\ref{sec:computation} for details on the computation.}
\label{fig:megafan}
\end{figure*}

In Refs.~\cite{Finney2022} and \cite{Wang2023}, we presented magnetotransport measurements and single-particle calculations for a TBG sample twisted to near 1.37\textdegree, where the miniband width is large enough (near 100 meV) to suppress correlated electronic states near zero magnetic field~\cite{Cao2018,Yankowitz2019,Finney2022}.  
In this work, we further explore that same sample, now at extremely high magnetic fields, demonstrating many $B$-induced strongly-correlated states. Most notably, at magnetic flux near half a magnetic flux quantum per moir\'e unit cell we observe plateaus in Hall resistance corresponding to $\pm8/3$ and $-8/5$, quantized to within a few tenths of a percent, along with possible $-4/3$ plateaus. The $\pm 8/3$ plateaus in particular extend over a strikingly broad density range, up to half an electron per moir\'e unit cell, rather than following a narrow \streda line. Integer quantized states have been known to extend over a broad density range, outcompeting fractional states that might be expected at those densities. This ``re-entrant quantum Hall effect'' has been explained as a consequence of a Wigner crystal coexisting with a filled Landau level~\cite{chen_competing_2019, das_sarma_perspectives_1996}, a scenario originally termed a partial Hall crystal~\cite{tesanovic_hall_1989}. In our present work, in contrast with prior reports, the integer states are {\em less} robust and broad than the $8/3$ fractional states.
In Sec.~\ref{sec:main_disc} we discuss several mechanisms that might be responsible for the novel and surprising extended FCI regions.

\section{Longitudinal transport}

In our sample, longitudinal resistivity as a function of carrier density and magnetic field between 14 and 28~T ranges from below our measurement floor of a few ohms within quantum Hall gaps to hundreds of kilohms at band edges and near charge neutrality (Fig.~\ref{fig:megafan}A). Here, we focus on specific contact pairs within this 20-terminal Hall bar that demonstrated the sharpest low-field transport in Ref.~\cite{Finney2022}. For our twist angle of 1.37\textdegree, $\Phi/\Phi_0=1$ occurs at nearly 45~T, so the field range in our high-field measurements corresponds to flux ratios between 0.31 and 0.62. We indicate the most prominent \streda lines, $(s,\ t)=(0,\, \pm4)$, which persist without closing down to near-zero field (Fig.~\ref{fig:tlcomp}).

\begin{figure*}[t!]
\centering
\includegraphics[width=18cm]{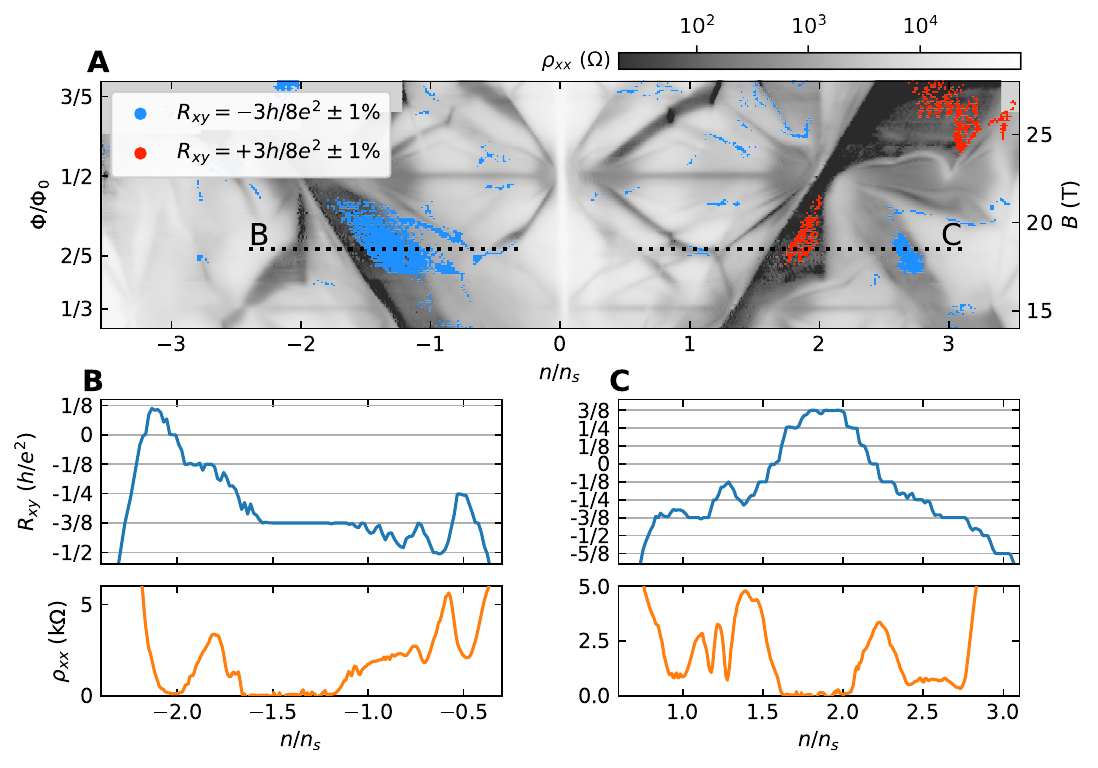}
\caption{\textbf{Fractional Chern insulating Hall plateaus.} (\textbf{A}) Longitudinal resistivity from Fig.~\ref{fig:megafan} replotted in gray-scale. Colored regions superimposed on the grey-scale plot mark where the Hall resistance for a neighboring contact pair is quantized to within 1\% of $-8/3$ (blue) or $+8/3$ (red). A clustering algorithm has been applied to remove spurious points (see Sec.~\ref{sec:clustering}). The abrupt loss of quantization at $\Phi/\Phi_0$ between $1/3$ and $2/5$ corresponds to a pause in the measurement of several hours. This, along with the noise above $1/2$, are described in Sec.~\ref{sec:noise}. (\textbf{B, C}) Hall resistance and longitudinal resistivity as a function of density at 18.5~T ($\Phi/\Phi_0=0.41$, black dashed lines in A).}
\label{fig:fci}
\end{figure*}

For comparison, we show the computed Wannier plot in the same field range (Fig.~\ref{fig:megafan}B). The computed density of states $dn/d\mu$ should show patterns similar to those seen in resistivity, but we do not directly calculate transport from our model as we did in~\cite{Wang2023} for lower magnetic fields. 
Our model, which incorporates uniaxial heterostrain, Zeeman effect, and electron-hole asymmetry, broadly matches our experimental observations of splitting and bending behavior of Landau levels emanating from half flux and $n/n_s=0$. The model predicts a number of prominent gaps that we discuss in Sec.~\ref{sec:main_computation} together with associated experimental measurements.

In the Supplement we describe other striking phenomena such as Brown-Zak oscillations, symmetry-broken Chern insulating \streda lines with $(s,t)=(\pm1/2,\pm3$), and correlated Hofstadter ferromagnets.

\section{Fractional Chern Insulators} \label{sec:main_fci}

In Fig.~\ref{fig:fci}A, we overlay in blue and red the regions where the Hall resistance is quantized to within 1\% of $-3h/8e^2$ and $+3h/8e^2$ respectively. 
We apply an intentionally-conservative clustering threshold algorithm so that almost all of the pockets marked with color represent well-quantized plateaus of $\pm 8/3$ rather than regions where the Hall resistance incidentally passes through quantized values. These fractional Hall plateaus coincide with low longitudinal resistivity (shown in gray). See Sec.~\ref{sec:clustering} for details and Fig.~\ref{fig:fci_raw} for unfiltered data.

Fig.~\ref{fig:fci} panels B and C show line cuts of $R_{xy}$ and $\rho_{xx}$ at 18.5~T. The aforementioned $\pm8/3$ plateaus are most prominent, but integer plateaus are visible, along with a small $-8/5$ plateau. For a discussion of the degree of quantization, other fractions, and the behavior in other contact pairs, see Secs.~\ref{sec:fci_quant}, ~\ref{sec:other_fracs}, and ~\ref{sec:otherpairs}, respectively.

\label{sec:computation}
\begin{figure*}[th]
\centering
\includegraphics[width=7.2in]{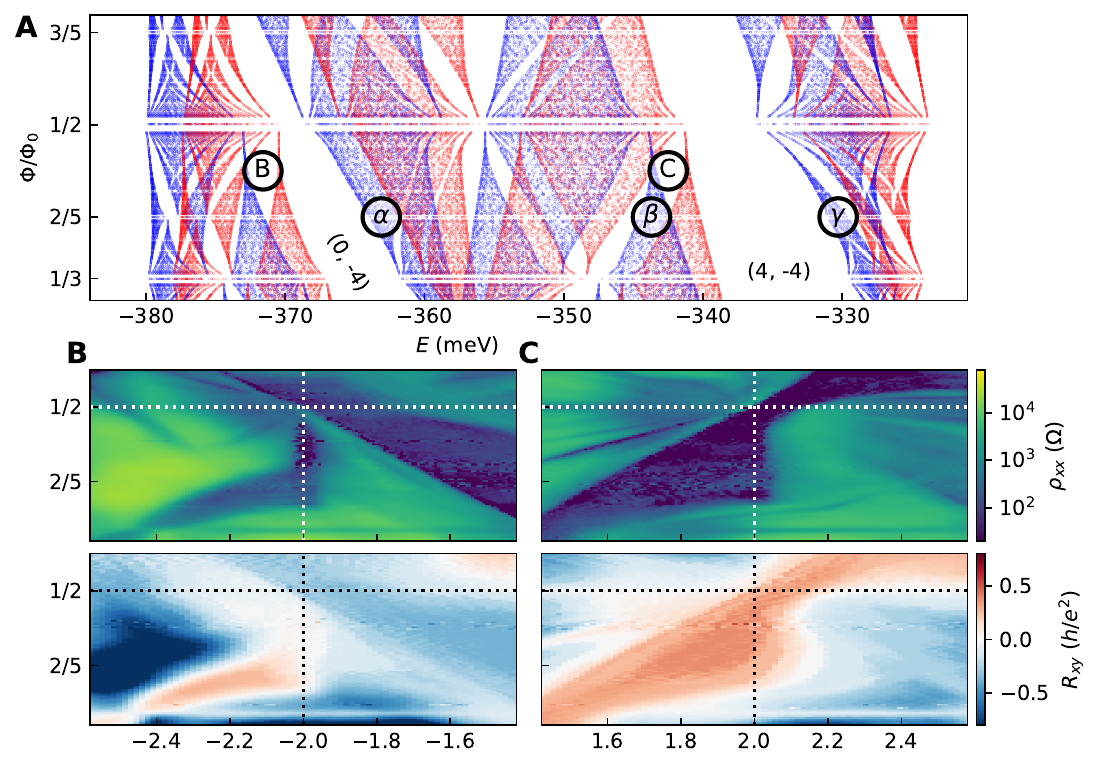}
\caption{\textbf{Computed Hofstadter spectrum and candidate quantum spin Hall state.} (\textbf{A}) Computed energy levels for K valley of strained TBG ($q\leq72$), as described in the text. Parameters: $\theta=1.35^\circ$, $0.24\%$ uniaxial heterostrain at 45\textdegree\ and $0.3\%$ biaxial heterostrain. Spin up electrons are shown in blue, and spin down electrons are shown in red. The gaps at $(s, t)=(0, -4)$ and $(4, -4)$ are labeled. $\alpha$, $\beta$, and $\gamma$ denote the Hofstadter subbands which host the largest pockets of $\pm8/3$ quantization shown in Fig.~\ref{fig:fci}. Two gaps $(s,t)=(0,\mp 2)$, marked B and C, are reminiscent of quantum spin Hall states. See Fig.~\ref{fig:computation_full} for the full range of fields.
(\textbf{B}-\textbf{C}) Dependence of the longitudinal resistivity (top panels) and Hall resistance (bottom panels) as a function of density and field centered on the two quantum spin Hall candidate gaps predicted from (A). In (B), below half flux we observe a strong suppression of longitudinal resistivity at fixed density $n/n_s=-2$ over a range of flux, coincident with a small plateau of zero Hall resistance. In (C), we observe low longitudinal resistivity along a vertical line, however the Hall resistance is not zero at the same densities.}
\label{fig:computation}
\end{figure*}

We find that regions where the Hall resistance is (fractionally) quantized typically coincide with suppressed longitudinal resistivity in the four adjacent contact pairs (see Fig.~\ref{fig:megafan} and Sec.~\ref{sec:otherpairs}). Though measurements from neighboring longitudinal probes vary subtly, neighboring Hall probes exhibit strikingly different behavior: those just 3 microns away from the main Hall pair we focus on in this manuscript exhibit poor quantization, even for integer quantum Hall states (Fig.~\ref{fig:waterfallish}), possibly due to mixing of the longitudinal resistivity into the Hall measurement because of spatial variation in twist angle~\footnote{The standard approach to remove the effect of mixing is to (anti)symmetrize the longitudinal resistivity (Hall resistance) as a function of magnetic field. Antisymmetrization could help reveal well-quantized plateaus in other Hall pairs. Given the time constraints of our measurement run at the National High Magnetic Field Lab, we were unable to acquire data for opposite field polarity.}. We therefore suspect that the main Hall pair of this manuscript contacts a region of unusually high moir\'e uniformity.
 
We defer a detailed discussion of potential explanations of the extended fractional phenomenology to Sec.~\ref{sec:main_disc}.

\section{Hofstadter Model} 
\label{sec:main_computation}

To understand the Hofstadter bands we would expect without interactions we perform bandstructure computations within a single-particle model. Our effective continuum model from Ref.~\cite{Wang2023}, which accounted for uniaxial heterostrain, can be extended into the Hofstadter regime~\cite{Wang2023_2}. Here we also incorporate biaxial heterostrain, lattice relaxation, and electron-hole asymmetry so that the locations of the three van Hove points near $B=0$ on the hole side of the charge neutrality point differ from those on the electron side~\cite{Vafek2023, Kang2023, kang_analytical_2025}, as is seen in the experiment~\cite{Finney2022, Wang2023}. We fully parameterize the $2\times2$ moir\'e deformation matrix as a function of twist angle, uniaxial heterostrain magnitude and direction, and biaxial heterostrain magnitude. 
We then search for structural parameters that yield bands with van Hove points at densities matching those identified at zero field in Ref.~\cite{Wang2023}.
We find a unique best fit---$0.24\pm0.02$\% uniaxial heterostrain at $45\pm4$\textdegree\ and $0.3\pm0.1$\% biaxial heterostrain---that places all six computed van Hove densities within our experimental bounds. See Sec.~\ref{supp:cont_model} for details.

Fig.~\ref{fig:computation}A shows the Hofstadter spectrum computed for this set of moir\'e parameters, over a range of flux filling corresponding to the magnetic field range of our measurements. We include a Zeeman splitting term $\Delta E=g\mu_B B$, with $g=2$ as expected for electron spins in graphene or other forms of carbon~\cite{prada_dirac_2021}. Spin-up electronic states (aligned with the external field) are shown in blue, and spin-down states in red. Computing the density of states d$n$/d$\mu$ of this spectrum facilitates comparison with our transport data (Fig.~\ref{fig:megafan}): the computed density of states qualitatively captures the extent in field and density of many well-formed gaps and subtler features in transport. However, not every feature in our non-interacting theory calculations matches experiment. 
Notably, the calculations show (4, -4) and (0, -4) as more prominent gaps than (0, 4) and (-4, 4).
In transport the (0, -4) \streda feature is indeed more prominent than (-4, 4). But the (0, 4) \streda feature is more prominent than (4, -4) (Fig.~\ref{fig:megafan}A), suggesting band renormalization effects beyond the single-particle picture ~\cite{Xie2020}.

Our model suggests that even without electron-electron interactions, gapped ground states with $t=0$ may exist in which the net spin of all filled bands is non-zero. We identify two such regimes, marked with B and C respectively in Fig.~\ref{fig:computation}A. In each case, just below half flux the Zeeman splitting exceeds the bandwidth of a subband separating two gaps associated with \streda lines that cross at half flux. The \streda
lines that cross at (B) are parameterized by (-4, 4) and (0, -4), and those at (C) by (0, 4) and (4, -4). The configuration at each of (B) and (C) is a quantum spin Hall insulator, with not one but two pairs of counter-propagating modes. We might therefore expect quantized longitudinal resistance $R_{xx} = h/{4e^2}$ and zero Hall resistance. The Hall resistance is indeed zero below half flux near density $n/n_s=-2$ (Fig.~\ref{fig:computation}B, lower), though not at $n/n_s=2$ (Fig.~\ref{fig:computation}C, lower). At $n/n_s=\pm2$, in longitudinal resistivity a narrow feature emanates downward from half flux(Fig.~\ref{fig:computation}B and C, upper). However, the longitudinal resistivity is not quantized as expected from our model, but is instead very low, often below our measurement floor of roughly 1~$\mathrm{\Omega}$ (in other contact pairs we see similar features but the minima are higher, see Fig.~\ref{fig:all_xx}). We see similar vertical features in $\rho_{xx}$ at quarter flux where the ($\pm$8, $\mp$8) and (0, $\mp$8) gaps intersect (see Fig.~\ref{fig:tlcomp}).
Given this unexpected behavior of longitudinal resistivity we cannot prove that some or all of these states in our experiment are QSH-like. If they indeed are QSH-like, our measurements suggest that the edge modes do not enter the ohmic contacts that serve as voltage probes, because of the narrowness of the arms connecting to the contacts and/or the multiple densities and thus fillings in series en route to the contacts. In this scenario, the voltage read out could be an average of the electrochemical potential of ballistically-propagating left-going and right-going modes, and thus could be almost identical at successive contacts.

\section{Landau level reset at half flux}
\label{sec:main_llreset}

\begin{figure*}[th]
\centering
\includegraphics[width=7.2in]{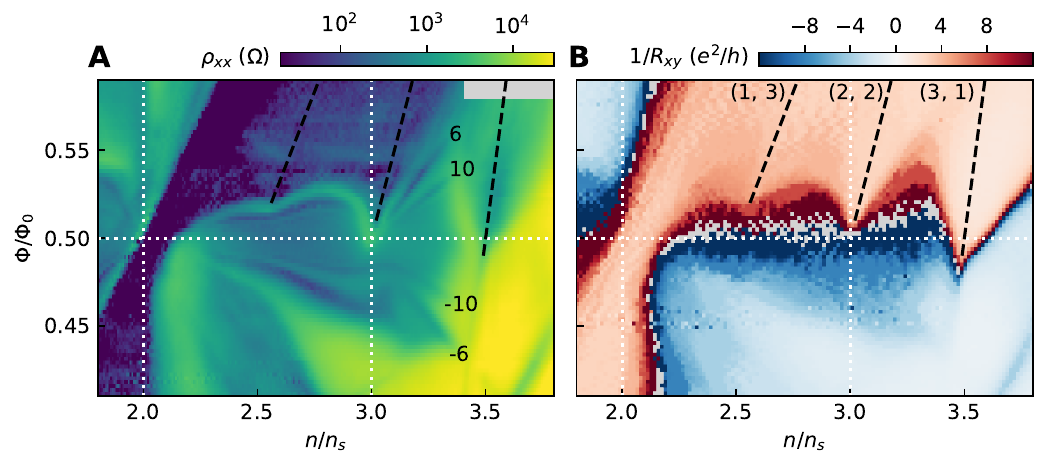}
\caption{\textbf{Landau level reset at half flux.} (\textbf{A}) Zoomed segment of Fig.~\ref{fig:megafan} showing the Landau levels bending and resetting at half flux. The numbers $\pm6$ and $\pm10$ are the slopes ($t$ values) of the faint lines $(3,\pm6)_{1/2}$ and $(3,\pm10)_{1/2}$ to their left. (\textbf{B}) Inverse Hall resistance showing that quantization of the Hall resistance tracks the bending dips in longitudinal resistivity.
Detail of all contact pairs in Figs.~\ref{fig:sharkbite}, \ref{fig:all_bendy}, and \ref{fig:bendy_xy}.}
\label{fig:bendymain}
\end{figure*}

We observe a bending Landau minifan emitting from $n/n_s=2,\Phi/\Phi_0=0.5$, where the $(0,\, 4)$ \streda line intersects half-flux (Fig.~\ref{fig:bendymain}). In longitudinal transport (panel A), the minifan pointing toward higher density and higher flux bends non-monotonically back toward half flux. Such bending is also visible in the minifan pointing toward lower flux, but in the following discussion we focus on the more prominent upward-pointing minifan. There is a small kink where the bending Landau minifan would have intersected the $(1, 3)$ \streda line near $n/n_s=2.5$, and a complete reset at $n/n_s=3$. In Hall measurements (panel B), we observe $h/4e^2$ and $h/8e^2$ integer quantum Hall plateaus
following the same bending behavior, including a third reset near $n/n_s=3.5$ (see also Fig.~\ref{fig:integers}).

This reset behavior is not the same bending behavior noted in Ref.~\cite{Finney2022} and cannot be understood in a non-interacting model of rigid bands. It is reminiscent of the Landau fan resets at integer filling in magic-angle twisted graphene structures~\cite{Cao2018_2,Yankowitz2019,Park2021,Wong2020,Tomarken2019,Park2021,Yu2022,Nuckolls2020,Pierce2021,Xie2021,Saito2021}. One possible explanation of this reset behavior in magic-angle samples is spontaneous isospin polarization. However, isospin polarization is inconsistent with our observation of a dip in longitudinal resistivity coincident with quantized Hall resistance following a nonlinear trajectory in the space of filling and flux; an incompressible Chern gap must follow the appropriate \streda relation.

For each of the four combinations of spin and valley, the two zero-field moir\'e minibands split into four magnetosubbands at half flux, reflecting a unit cell doubling in the Hofstadter spectrum.  
Because of this unit cell doubling, integer filling of the highest-energy magnetosubband (two valley-degenerate subbands split by Zeeman) corresponds to fillings $n/n_s=2.5$, $3$, $3.5$, and $4$ referenced to the original unit cell. 
At $n/n_s=3$, if the carriers within this band fully polarize into half of the available flavors, we expect a total Chern number of all filled bands of 2.
One can see this from the fact that this gap must be the same as the (2, 2) Hofstadter subband ferromagnet state emanating from zero field~\cite{Saito2021,Wang2024}.
Landau levels emanating from half flux carry an extra two-fold degeneracy (see Sec.~\ref{sec:resetzoom}), therefore we would expect a four-fold degenerate fan with an offset of 2 ($t=\pm 2,6,10, \ldots$) emanating from $n/n_s=3$.
In our measurements, the Landau levels emanating from the reset point are indeed fourfold degenerate, with $(3,\pm6)_{1/2}$ and $(3,\pm10)_{1/2}$ being the only visible gaps. Here, we use the subscript $1/2$ to indicate we are using $s$ to reference a density offset at half of a flux quantum per moir\'e unit cell. 

The resets do not all appear at exactly half flux. Rather, the reset seen in both longitudinal and Hall measurements near $n/n_s=2.5$ is just above half flux, and the reset seen in Hall at $n/n_s=3.5$ is just below half flux. Each reset aligns with an extrapolation of a Hofstadter ferromagnetic state---$(1, 3)$, $(2, 2)$, and $(3, 1)$, respectively---from zero field (dashed lines).

\section{Discussion of possible mechanisms for extended FCI} \label{sec:main_disc}

We turn now to the $8/3$ fractional state which surprisingly persists over a larger range of density than any integer state. 
A linecut at $18.5$~T, near 2/5 flux (Fig.~\ref{fig:fci}B, C), traverses the three largest pockets of quantization in the Landau fan diagram, centered at $n/n_s= -1.3, 1.9$, and $2.6$. These occur upon doping into the next band above the Hofstadter gaps $(0, -4)$, $(0, +4)$, and $(+4, -4)$, respectively. Assuming the total Chern number $\pm 8/3$ extracted from measured Hall resistance is the sum of the Chern number $\Delta t$ of the FCI and that of the gapped integer state below, the Chern numbers of the three FCI states would be $+4/3=-8/3-(-4)$, $-4/3=8/3-4$, and again $+4/3=-8/3-(-4)$, respectively. In Fig.~\ref{fig:computation}A, within our single-particle model we denote the parent spin-polarized Hofstadter subbands at 2/5 flux that we believe host the FCI states $\alpha$, $\beta$, and $\gamma$, respectively. If interactions do not generate substantial mixing of bands, in each case we are doping into a spin-polarized (all blue in the figure) but valley-degenerate band. In this scenario, the experimentally-extracted 4/3 Chern number represents two copies of a $|\Delta t| = 2/3$ FCI. Alternatively, if valley degeneracy is spontaneously broken by interactions the observed quantization could reflect a non-degenerate $|\Delta t| = 4/3$ FCI. We cannot rule out the possibility that it is four copies of a 1/3 FCI, but that would require recovery of spin degeneracy in the face of a Zeeman energy comparable to subband widths.

Unlike ordinary Landau levels, the three subbands we postulate as parents for the FCIs have finite bandwidth, even without disorder. Subband $\gamma$ is the narrowest at $\sim 1$ meV. The others have width $\sim 2$ meV at 2/5 flux, and the associated FCIs extend over a range of flux for which the bands broaden further and even overlap with nearby subbands, factors we would ordinarily expect to disfavor formation of FCI states.

Subband $\alpha$ and its corresponding Hall plateau are particularly enigmatic. We observe an enormous region of quantization ($\Delta n/n_s\sim0.5$), larger than any other plateau---integer or fractional---between $0$ and $28$ T (see Fig.~\ref{fig:integers}).
An incompressible gap is expected to follow a well-defined \streda relation with slope equal to the total Chern number of the occupied bands.
However, the breadth of this plateau stymies assigning a specific ($s$, $t$). 
To retain quantization as carrier density is tuned, the excess carriers must go into localized states. What states are we doping into here? The obvious culprit would be localized states from disorder.  
Though we cannot explicitly rule out this scenario, it is unlikely that localized states from disorder would specifically extend these $\pm$8/3 fractional gaps so much more dramatically than other gaps in the spectrum. In Sec.~\ref{sec:domain_wall} and Sec.~\ref{sec:spindep}, we discuss the effects of domain walls and spin-dependent transport, respectively. We consider both scenarios possible but unlikely as explanations of our data.

A more likely reason for these fractional plateaus to be so large is the formation of a fractional partial Hall crystal: a Wigner crystal in a FCI background. In this scenario, upon doping with respect to primitive filling of the FCI gap, the strongly-interacting dilute system of quasiparticles forms a Wigner crystal. As this crystal is topologically trivial, insulating, and compressible, it does not contribute to transport, so quantized Hall resistance is retained or restored at the value associated with the FCI state~\cite{tesanovic_hall_1989}.

This mechanism has been invoked to explain reentrant (fractional) quantum Hall in 2D electron gases~\cite{chen_competing_2019, das_sarma_perspectives_1996, shingla_highly_2023}, but for the present system we have two complicating factors. First, the presence of a moir\'e potential might be expected to favor forming a Wigner crystal specifically at rational fractional fillings of the moir\'e. We instead observe sustained quantization over a range of density that would encompass many simple fractions.
Second, this plateau occurs in a regime of two valley-degenerate Hofstadter subbands, with bandwidth comparable to Zeeman splitting. There could thus be more than one way for the multiple flavors to combine into a fractional state, reminiscent of $\nu=2/3$ fractional quantum Hall in a seminconductor double quantum well~\cite{mcdonald_topological_1996}.
Understanding the nature of the extended FCIs and the conditions for a subband to fractionalize warrant further experimental and theoretical investigation. If the electronic solid picture we suggest is correct, disentangling the precise nature of the solid, including whether it is partially or entirely composed of a second flavor of the composite fermions themselves, will be an important challenge.

\textbf{Acknowledgments}
We would like to thank Steve Kivelson, Julian May-Mann, Trithep Devakul, Yves Kwan, Tomohiro Soejima, Patrick Ledwith, Ben Feldman, Matt Yankowitz, Dave Cobden, Xiaodong Xu, Wei Pan, Jonah Herzog-Arbeitman, Nisarga Paul, and Sayak Bhattacharjee for fruitful discussions. 
Sample preparation, measurements, and analysis were supported by the US Department of Energy, Office of Science, Basic Energy Sciences, Materials Sciences and Engineering Division, under Contract DE-AC02-76SF00515. 
Device fabrication was performed at the Stanford Nano Shared Facilities which is supported by the NSF under award ECCS-2026822. 
D.G.-G. acknowledges support for supplies from the Ross M. Brown Family Foundation and from the Gordon and Betty Moore Foundation’s Emergent Phenomena in Quantum Systems (EPiQS) Initiative through grant GBMF9460.
X.W. acknowledges financial support from National MagLab through Dirac fellowship, which is funded by the National Science Foundation (Grant No. DMR-2128556) and the state of Florida.
O.V. was funded in part by the Gordon and Betty Moore Foundation’s EPiQS Initiative Grant GBMF11070 and acknowledges support from the National High Magnetic Field Laboratory funded by the National Science Foundation (Grant No. DMR-2128556) and the State of Florida.
K.W. and T.T. acknowledge support from JSPS KAKENHI (Grant Numbers 19H05790, 20H00354 and 21H05233). 
Part of this work was performed at the Stanford Nano Shared Facilities (SNSF), supported by the National Science Foundation under award ECCS-2026822.
A portion of this work was performed at the National High Magnetic Field Laboratory, which is supported by the National Science Foundation Cooperative Agreement No. DMR-1644779 and the State of Florida.

\textbf{Author Contributions}
J.F. and A.S. devised the project, performed data analysis, and wrote the manuscript with input from all authors.
J.F. and L.K.R. fabricated the devices and performed measurements. 
J.K. X.W. and O.V. performed bandstructure calculations.
K.W. and T.T. supplied the boron nitride crystals.
M.A.K. and D.G-G. supervised the project.

\textbf{Competing interests}
The authors declare no competing interests.

\bibliographystyle{naturemag}
\bibliography{bib, aaron_references}

\begin{thebibliography}{10}
\expandafter\ifx\csname url\endcsname\relax
  \def\url#1{\texttt{#1}}\fi
\expandafter\ifx\csname urlprefix\endcsname\relax\def\urlprefix{URL }\fi
\providecommand{\bibinfo}[2]{#2}
\providecommand{\eprint}[2][]{\url{#2}}

\bibitem{Cao2018}
\bibinfo{author}{Cao, Y.} \emph{et~al.}
\newblock \bibinfo{title}{Correlated insulator behaviour at half-filling in
  magic-angle graphene superlattices}.
\newblock \emph{\bibinfo{journal}{Nature}} \textbf{\bibinfo{volume}{556}},
  \bibinfo{pages}{80–84} (\bibinfo{year}{2018}).
\newblock \urlprefix\url{http://dx.doi.org/10.1038/nature26154}.

\bibitem{Cao2018_2}
\bibinfo{author}{Cao, Y.} \emph{et~al.}
\newblock \bibinfo{title}{Unconventional superconductivity in magic-angle
  graphene superlattices}.
\newblock \emph{\bibinfo{journal}{Nature}} \textbf{\bibinfo{volume}{556}},
  \bibinfo{pages}{43–50} (\bibinfo{year}{2018}).
\newblock \urlprefix\url{http://dx.doi.org/10.1038/nature26160}.

\bibitem{Yankowitz2019}
\bibinfo{author}{Yankowitz, M.} \emph{et~al.}
\newblock \bibinfo{title}{Tuning superconductivity in twisted bilayer
  graphene}.
\newblock \emph{\bibinfo{journal}{Science}} \textbf{\bibinfo{volume}{363}},
  \bibinfo{pages}{1059–1064} (\bibinfo{year}{2019}).
\newblock \urlprefix\url{http://dx.doi.org/10.1126/science.aav1910}.

\bibitem{Sharpe2019}
\bibinfo{author}{Sharpe, A.~L.} \emph{et~al.}
\newblock \bibinfo{title}{Emergent ferromagnetism near three-quarters filling
  in twisted bilayer graphene}.
\newblock \emph{\bibinfo{journal}{Science}} \textbf{\bibinfo{volume}{365}},
  \bibinfo{pages}{605–608} (\bibinfo{year}{2019}).
\newblock \urlprefix\url{http://dx.doi.org/10.1126/science.aaw3780}.

\bibitem{Andrei2020}
\bibinfo{author}{Andrei, E.~Y.} \& \bibinfo{author}{MacDonald, A.~H.}
\newblock \bibinfo{title}{Graphene bilayers with a twist}.
\newblock \emph{\bibinfo{journal}{Nature Materials}}
  \textbf{\bibinfo{volume}{19}}, \bibinfo{pages}{1265–1275}
  (\bibinfo{year}{2020}).
\newblock \urlprefix\url{http://dx.doi.org/10.1038/s41563-020-00840-0}.

\bibitem{Balents2020}
\bibinfo{author}{Balents, L.}, \bibinfo{author}{Dean, C.~R.},
  \bibinfo{author}{Efetov, D.~K.} \& \bibinfo{author}{Young, A.~F.}
\newblock \bibinfo{title}{Superconductivity and strong correlations in moiré
  flat bands}.
\newblock \emph{\bibinfo{journal}{Nature Physics}}
  \textbf{\bibinfo{volume}{16}}, \bibinfo{pages}{725–733}
  (\bibinfo{year}{2020}).
\newblock \urlprefix\url{http://dx.doi.org/10.1038/s41567-020-0906-9}.

\bibitem{Pierce2021}
\bibinfo{author}{Pierce, A.~T.} \emph{et~al.}
\newblock \bibinfo{title}{Unconventional sequence of correlated chern
  insulators in magic-angle twisted bilayer graphene}.
\newblock \emph{\bibinfo{journal}{Nature Physics}}
  \textbf{\bibinfo{volume}{17}}, \bibinfo{pages}{1210–1215}
  (\bibinfo{year}{2021}).
\newblock \urlprefix\url{http://dx.doi.org/10.1038/s41567-021-01347-4}.

\bibitem{Xie2021}
\bibinfo{author}{Xie, Y.} \emph{et~al.}
\newblock \bibinfo{title}{Fractional chern insulators in magic-angle twisted
  bilayer graphene}.
\newblock \emph{\bibinfo{journal}{Nature}} \textbf{\bibinfo{volume}{600}},
  \bibinfo{pages}{439–443} (\bibinfo{year}{2021}).
\newblock \urlprefix\url{http://dx.doi.org/10.1038/s41586-021-04002-3}.

\bibitem{Saito2021}
\bibinfo{author}{Saito, Y.} \emph{et~al.}
\newblock \bibinfo{title}{Hofstadter subband ferromagnetism and symmetry-broken
  chern insulators in twisted bilayer graphene}.
\newblock \emph{\bibinfo{journal}{Nature Physics}}
  \textbf{\bibinfo{volume}{17}}, \bibinfo{pages}{478–481}
  (\bibinfo{year}{2021}).
\newblock \urlprefix\url{http://dx.doi.org/10.1038/s41567-020-01129-4}.

\bibitem{Liu2021}
\bibinfo{author}{Liu, X.} \emph{et~al.}
\newblock \bibinfo{title}{Tuning electron correlation in magic-angle twisted
  bilayer graphene using coulomb screening}.
\newblock \emph{\bibinfo{journal}{Science}} \textbf{\bibinfo{volume}{371}},
  \bibinfo{pages}{1261–1265} (\bibinfo{year}{2021}).
\newblock \urlprefix\url{http://dx.doi.org/10.1126/science.abb8754}.

\bibitem{Lu2021}
\bibinfo{author}{Lu, X.} \emph{et~al.}
\newblock \bibinfo{title}{Multiple flat bands and topological hofstadter
  butterfly in twisted bilayer graphene close to the second magic angle}.
\newblock \emph{\bibinfo{journal}{Proceedings of the National Academy of
  Sciences}} \textbf{\bibinfo{volume}{118}} (\bibinfo{year}{2021}).
\newblock \urlprefix\url{http://dx.doi.org/10.1073/pnas.2100006118}.

\bibitem{Stepanov2021}
\bibinfo{author}{Stepanov, P.} \emph{et~al.}
\newblock \bibinfo{title}{Competing zero-field chern insulators in
  superconducting twisted bilayer graphene}.
\newblock \emph{\bibinfo{journal}{Physical Review Letters}}
  \textbf{\bibinfo{volume}{127}} (\bibinfo{year}{2021}).
\newblock \urlprefix\url{http://dx.doi.org/10.1103/PhysRevLett.127.197701}.

\bibitem{Hofstadter1976}
\bibinfo{author}{Hofstadter, D.~R.}
\newblock \bibinfo{title}{Energy levels and wave functions of bloch electrons
  in rational and irrational magnetic fields}.
\newblock \emph{\bibinfo{journal}{Physical Review B}}
  \textbf{\bibinfo{volume}{14}}, \bibinfo{pages}{2239–2249}
  (\bibinfo{year}{1976}).
\newblock \urlprefix\url{http://dx.doi.org/10.1103/PhysRevB.14.2239}.

\bibitem{Bistritzer2011_2}
\bibinfo{author}{Bistritzer, R.} \& \bibinfo{author}{MacDonald, A.~H.}
\newblock \bibinfo{title}{Moiré butterflies in twisted bilayer graphene}.
\newblock \emph{\bibinfo{journal}{Physical Review B}}
  \textbf{\bibinfo{volume}{84}} (\bibinfo{year}{2011}).
\newblock \urlprefix\url{http://dx.doi.org/10.1103/PhysRevB.84.035440}.

\bibitem{Dean2013}
\bibinfo{author}{Dean, C.~R.} \emph{et~al.}
\newblock \bibinfo{title}{Hofstadter’s butterfly and the fractal quantum hall
  effect in moiré superlattices}.
\newblock \emph{\bibinfo{journal}{Nature}} \textbf{\bibinfo{volume}{497}},
  \bibinfo{pages}{598–602} (\bibinfo{year}{2013}).
\newblock \urlprefix\url{http://dx.doi.org/10.1038/nature12186}.

\bibitem{Hunt2013}
\bibinfo{author}{Hunt, B.} \emph{et~al.}
\newblock \bibinfo{title}{Massive dirac fermions and hofstadter butterfly in a
  van der waals heterostructure}.
\newblock \emph{\bibinfo{journal}{Science}} \textbf{\bibinfo{volume}{340}},
  \bibinfo{pages}{1427–1430} (\bibinfo{year}{2013}).
\newblock \urlprefix\url{http://dx.doi.org/10.1126/science.1237240}.

\bibitem{Ponomarenko2013}
\bibinfo{author}{Ponomarenko, L.~A.} \emph{et~al.}
\newblock \bibinfo{title}{Cloning of dirac fermions in graphene superlattices}.
\newblock \emph{\bibinfo{journal}{Nature}} \textbf{\bibinfo{volume}{497}},
  \bibinfo{pages}{594–597} (\bibinfo{year}{2013}).
\newblock \urlprefix\url{http://dx.doi.org/10.1038/nature12187}.

\bibitem{Wang2015}
\bibinfo{author}{Wang, L.} \emph{et~al.}
\newblock \bibinfo{title}{Evidence for a fractional fractal quantum hall effect
  in graphene superlattices}.
\newblock \emph{\bibinfo{journal}{Science}} \textbf{\bibinfo{volume}{350}},
  \bibinfo{pages}{1231–1234} (\bibinfo{year}{2015}).
\newblock \urlprefix\url{http://dx.doi.org/10.1126/science.aad2102}.

\bibitem{Spanton2018}
\bibinfo{author}{Spanton, E.~M.} \emph{et~al.}
\newblock \bibinfo{title}{Observation of fractional chern insulators in a van
  der waals heterostructure}.
\newblock \emph{\bibinfo{journal}{Science}} \textbf{\bibinfo{volume}{360}},
  \bibinfo{pages}{62–66} (\bibinfo{year}{2018}).
\newblock \urlprefix\url{http://dx.doi.org/10.1126/science.aan8458}.

\bibitem{das_observation_2022}
\bibinfo{author}{Das, I.} \emph{et~al.}
\newblock \bibinfo{title}{Observation of {Reentrant} {Correlated} {Insulators}
  and {Interaction}-{Driven} {Fermi}-{Surface} {Reconstructions} at {One}
  {Magnetic} {Flux} {Quantum} per {Moir}{\textbackslash}'e {Unit} {Cell} in
  {Magic}-{Angle} {Twisted} {Bilayer} {Graphene}}.
\newblock \emph{\bibinfo{journal}{Physical Review Letters}}
  \textbf{\bibinfo{volume}{128}}, \bibinfo{pages}{217701}
  (\bibinfo{year}{2022}).
\newblock
  \urlprefix\url{https://link.aps.org/doi/10.1103/PhysRevLett.128.217701}.
\newblock \bibinfo{note}{Publisher: American Physical Society}.

\bibitem{Wang2022}
\bibinfo{author}{Wang, X.} \& \bibinfo{author}{Vafek, O.}
\newblock \bibinfo{title}{Narrow bands in magnetic field and strong-coupling
  hofstadter spectra}.
\newblock \emph{\bibinfo{journal}{Physical Review B}}
  \textbf{\bibinfo{volume}{106}} (\bibinfo{year}{2022}).
\newblock \urlprefix\url{https://doi.org/10.1103/physrevb.106.l121111}.

\bibitem{Yu2022}
\bibinfo{author}{Yu, J.} \emph{et~al.}
\newblock \bibinfo{title}{Correlated hofstadter spectrum and flavour phase
  diagram in magic-angle twisted bilayer graphene}.
\newblock \emph{\bibinfo{journal}{Nature Physics}}
  \textbf{\bibinfo{volume}{18}}, \bibinfo{pages}{825–831}
  (\bibinfo{year}{2022}).
\newblock \urlprefix\url{http://dx.doi.org/10.1038/s41567-022-01589-w}.

\bibitem{herzog-arbeitman_reentrant_2022}
\bibinfo{author}{Herzog-Arbeitman, J.}, \bibinfo{author}{Chew, A.},
  \bibinfo{author}{Efetov, D.~K.} \& \bibinfo{author}{Bernevig, B.~A.}
\newblock \bibinfo{title}{Reentrant {Correlated} {Insulators} in {Twisted}
  {Bilayer} {Graphene} at 25 {T}
  (\$2{\textbackslash}ensuremath\{{\textbackslash}pi\}\$ {Flux})}.
\newblock \emph{\bibinfo{journal}{Physical Review Letters}}
  \textbf{\bibinfo{volume}{129}}, \bibinfo{pages}{076401}
  (\bibinfo{year}{2022}).
\newblock
  \urlprefix\url{https://link.aps.org/doi/10.1103/PhysRevLett.129.076401}.

\bibitem{Wang2024}
\bibinfo{author}{Wang, X.} \& \bibinfo{author}{Vafek, O.}
\newblock \bibinfo{title}{Theory of correlated chern insulators in twisted
  bilayer graphene}.
\newblock \emph{\bibinfo{journal}{Physical Review X}}
  \textbf{\bibinfo{volume}{14}} (\bibinfo{year}{2024}).
\newblock \urlprefix\url{http://dx.doi.org/10.1103/PhysRevX.14.021042}.

\bibitem{Wannier1978}
\bibinfo{author}{Wannier, G.~H.}
\newblock \bibinfo{title}{A result not dependent on rationality for bloch
  electrons in a magnetic field}.
\newblock \emph{\bibinfo{journal}{physica status solidi (b)}}
  \textbf{\bibinfo{volume}{88}}, \bibinfo{pages}{757–765}
  (\bibinfo{year}{1978}).
\newblock \urlprefix\url{http://dx.doi.org/10.1002/pssb.2220880243}.

\bibitem{Thouless1982}
\bibinfo{author}{Thouless, D.~J.}, \bibinfo{author}{Kohmoto, M.},
  \bibinfo{author}{Nightingale, M.~P.} \& \bibinfo{author}{den Nijs, M.}
\newblock \bibinfo{title}{Quantized hall conductance in a two-dimensional
  periodic potential}.
\newblock \emph{\bibinfo{journal}{Physical Review Letters}}
  \textbf{\bibinfo{volume}{49}}, \bibinfo{pages}{405–408}
  (\bibinfo{year}{1982}).
\newblock \urlprefix\url{http://dx.doi.org/10.1103/PhysRevLett.49.405}.

\bibitem{Young2012}
\bibinfo{author}{Young, A.~F.} \emph{et~al.}
\newblock \bibinfo{title}{Spin and valley quantum hall ferromagnetism in
  graphene}.
\newblock \emph{\bibinfo{journal}{Nature Physics}}
  \textbf{\bibinfo{volume}{8}}, \bibinfo{pages}{550–556}
  (\bibinfo{year}{2012}).
\newblock \urlprefix\url{http://dx.doi.org/10.1038/nphys2307}.

\bibitem{kol_fractional_1993}
\bibinfo{author}{Kol, A.} \& \bibinfo{author}{Read, N.}
\newblock \bibinfo{title}{Fractional quantum {Hall} effect in a periodic
  potential}.
\newblock \emph{\bibinfo{journal}{Physical Review B}}
  \textbf{\bibinfo{volume}{48}}, \bibinfo{pages}{8890--8898}
  (\bibinfo{year}{1993}).
\newblock \urlprefix\url{https://link.aps.org/doi/10.1103/PhysRevB.48.8890}.
\newblock \bibinfo{note}{Publisher: American Physical Society}.

\bibitem{neupert_fractional_quantum_2011}
\bibinfo{author}{Neupert, T.}, \bibinfo{author}{Santos, L.},
  \bibinfo{author}{Chamon, C.} \& \bibinfo{author}{Mudry, C.}
\newblock \bibinfo{title}{Fractional {Quantum} {Hall} {States} at {Zero}
  {Magnetic} {Field}}.
\newblock \emph{\bibinfo{journal}{Physical Review Letters}}
  \textbf{\bibinfo{volume}{106}}, \bibinfo{pages}{236804}
  (\bibinfo{year}{2011}).
\newblock
  \urlprefix\url{https://link.aps.org/doi/10.1103/PhysRevLett.106.236804}.
\newblock \bibinfo{note}{Publisher: American Physical Society}.

\bibitem{neupert_fractional_2011}
\bibinfo{author}{Neupert, T.}, \bibinfo{author}{Santos, L.},
  \bibinfo{author}{Ryu, S.}, \bibinfo{author}{Chamon, C.} \&
  \bibinfo{author}{Mudry, C.}
\newblock \bibinfo{title}{Fractional topological liquids with time-reversal
  symmetry and their lattice realization}.
\newblock \emph{\bibinfo{journal}{Physical Review B}}
  \textbf{\bibinfo{volume}{84}}, \bibinfo{pages}{165107}
  (\bibinfo{year}{2011}).
\newblock \urlprefix\url{https://link.aps.org/doi/10.1103/PhysRevB.84.165107}.
\newblock \bibinfo{note}{Publisher: American Physical Society}.

\bibitem{regnault_fractional_2011}
\bibinfo{author}{Regnault, N.} \& \bibinfo{author}{Bernevig, B.~A.}
\newblock \bibinfo{title}{Fractional {Chern} {Insulator}}.
\newblock \emph{\bibinfo{journal}{Physical Review X}}
  \textbf{\bibinfo{volume}{1}}, \bibinfo{pages}{021014} (\bibinfo{year}{2011}).
\newblock \urlprefix\url{https://link.aps.org/doi/10.1103/PhysRevX.1.021014}.
\newblock \bibinfo{note}{Publisher: American Physical Society}.

\bibitem{santos_time-reversal_2011}
\bibinfo{author}{Santos, L.}, \bibinfo{author}{Neupert, T.},
  \bibinfo{author}{Ryu, S.}, \bibinfo{author}{Chamon, C.} \&
  \bibinfo{author}{Mudry, C.}
\newblock \bibinfo{title}{Time-reversal symmetric hierarchy of fractional
  incompressible liquids}.
\newblock \emph{\bibinfo{journal}{Physical Review B}}
  \textbf{\bibinfo{volume}{84}}, \bibinfo{pages}{165138}
  (\bibinfo{year}{2011}).
\newblock \urlprefix\url{https://link.aps.org/doi/10.1103/PhysRevB.84.165138}.
\newblock \bibinfo{note}{Publisher: American Physical Society}.

\bibitem{sheng_fractional_2011}
\bibinfo{author}{Sheng, D.~N.}, \bibinfo{author}{Gu, Z.-C.},
  \bibinfo{author}{Sun, K.} \& \bibinfo{author}{Sheng, L.}
\newblock \bibinfo{title}{Fractional quantum {Hall} effect in the absence of
  {Landau} levels}.
\newblock \emph{\bibinfo{journal}{Nature Communications}}
  \textbf{\bibinfo{volume}{2}}, \bibinfo{pages}{389} (\bibinfo{year}{2011}).
\newblock \urlprefix\url{https://www.nature.com/articles/ncomms1380}.
\newblock \bibinfo{note}{Publisher: Nature Publishing Group}.

\bibitem{sun_nearly_2011}
\bibinfo{author}{Sun, K.}, \bibinfo{author}{Gu, Z.}, \bibinfo{author}{Katsura,
  H.} \& \bibinfo{author}{Das~Sarma, S.}
\newblock \bibinfo{title}{Nearly {Flatbands} with {Nontrivial} {Topology}}.
\newblock \emph{\bibinfo{journal}{Physical Review Letters}}
  \textbf{\bibinfo{volume}{106}}, \bibinfo{pages}{236803}
  (\bibinfo{year}{2011}).
\newblock
  \urlprefix\url{https://link.aps.org/doi/10.1103/PhysRevLett.106.236803}.
\newblock \bibinfo{note}{Publisher: American Physical Society}.

\bibitem{tang_high-temperature_2011}
\bibinfo{author}{Tang, E.}, \bibinfo{author}{Mei, J.-W.} \&
  \bibinfo{author}{Wen, X.-G.}
\newblock \bibinfo{title}{High-{Temperature} {Fractional} {Quantum} {Hall}
  {States}}.
\newblock \emph{\bibinfo{journal}{Physical Review Letters}}
  \textbf{\bibinfo{volume}{106}}, \bibinfo{pages}{236802}
  (\bibinfo{year}{2011}).
\newblock
  \urlprefix\url{https://link.aps.org/doi/10.1103/PhysRevLett.106.236802}.
\newblock \bibinfo{note}{Publisher: American Physical Society}.

\bibitem{Laughlin1983}
\bibinfo{author}{Laughlin, R.~B.}
\newblock \bibinfo{title}{Anomalous quantum hall effect: An incompressible
  quantum fluid with fractionally charged excitations}.
\newblock \emph{\bibinfo{journal}{Physical Review Letters}}
  \textbf{\bibinfo{volume}{50}}, \bibinfo{pages}{1395–1398}
  (\bibinfo{year}{1983}).
\newblock \urlprefix\url{http://dx.doi.org/10.1103/PhysRevLett.50.1395}.

\bibitem{Cai2023}
\bibinfo{author}{Cai, J.} \emph{et~al.}
\newblock \bibinfo{title}{Signatures of fractional quantum anomalous hall
  states in twisted mote2}.
\newblock \emph{\bibinfo{journal}{Nature}} \textbf{\bibinfo{volume}{622}},
  \bibinfo{pages}{63–68} (\bibinfo{year}{2023}).
\newblock \urlprefix\url{http://dx.doi.org/10.1038/s41586-023-06289-w}.

\bibitem{Zeng2023}
\bibinfo{author}{Zeng, Y.} \emph{et~al.}
\newblock \bibinfo{title}{Thermodynamic evidence of fractional chern insulator
  in moiré mote2}.
\newblock \emph{\bibinfo{journal}{Nature}} \textbf{\bibinfo{volume}{622}},
  \bibinfo{pages}{69–73} (\bibinfo{year}{2023}).
\newblock \urlprefix\url{http://dx.doi.org/10.1038/s41586-023-06452-3}.

\bibitem{Park2023}
\bibinfo{author}{Park, H.} \emph{et~al.}
\newblock \bibinfo{title}{Observation of fractionally quantized anomalous hall
  effect}.
\newblock \emph{\bibinfo{journal}{Nature}} \textbf{\bibinfo{volume}{622}},
  \bibinfo{pages}{74–79} (\bibinfo{year}{2023}).
\newblock \urlprefix\url{http://dx.doi.org/10.1038/s41586-023-06536-0}.

\bibitem{Lu2024}
\bibinfo{author}{Lu, Z.} \emph{et~al.}
\newblock \bibinfo{title}{Fractional quantum anomalous hall effect in
  multilayer graphene}.
\newblock \emph{\bibinfo{journal}{Nature}} \textbf{\bibinfo{volume}{626}},
  \bibinfo{pages}{759–764} (\bibinfo{year}{2024}).
\newblock \urlprefix\url{http://dx.doi.org/10.1038/s41586-023-07010-7}.

\bibitem{luExtendedQuantumAnomalous2025}
\bibinfo{author}{Lu, Z.} \emph{et~al.}
\newblock \bibinfo{title}{Extended quantum anomalous {{Hall}} states in
  graphene/{{hBN}} moir{\'e} superlattices}.
\newblock \emph{\bibinfo{journal}{Nature}} \textbf{\bibinfo{volume}{637}},
  \bibinfo{pages}{1090--1095} (\bibinfo{year}{2025}).
\newblock \urlprefix\url{https://doi.org/10.1038/s41586-024-08470-1}.

\bibitem{Parker2021}
\bibinfo{author}{Parker, D.~E.}, \bibinfo{author}{Soejima, T.},
  \bibinfo{author}{Hauschild, J.}, \bibinfo{author}{Zaletel, M.~P.} \&
  \bibinfo{author}{Bultinck, N.}
\newblock \bibinfo{title}{Strain-induced quantum phase transitions in
  magic-angle graphene}.
\newblock \emph{\bibinfo{journal}{Physical Review Letters}}
  \textbf{\bibinfo{volume}{127}} (\bibinfo{year}{2021}).
\newblock \urlprefix\url{http://dx.doi.org/10.1103/PhysRevLett.127.027601}.

\bibitem{he_strongly_2024}
\bibinfo{author}{He, M.} \emph{et~al.}
\newblock \bibinfo{title}{Strongly interacting {Hofstadter} states in
  magic-angle twisted bilayer graphene} (\bibinfo{year}{2024}).
\newblock \urlprefix\url{https://arxiv.org/abs/2408.01599v1}.

\bibitem{Finney2022}
\bibinfo{author}{Finney, J.} \emph{et~al.}
\newblock \bibinfo{title}{Unusual magnetotransport in twisted bilayer
  graphene}.
\newblock \emph{\bibinfo{journal}{Proceedings of the National Academy of
  Sciences}} \textbf{\bibinfo{volume}{119}} (\bibinfo{year}{2022}).
\newblock \urlprefix\url{https://doi.org/10.1073/pnas.2118482119}.

\bibitem{Wang2023}
\bibinfo{author}{Wang, X.} \emph{et~al.}
\newblock \bibinfo{title}{Unusual magnetotransport in twisted bilayer graphene
  from strain-induced open fermi surfaces}.
\newblock \emph{\bibinfo{journal}{Proceedings of the National Academy of
  Sciences}} \textbf{\bibinfo{volume}{120}} (\bibinfo{year}{2023}).
\newblock \urlprefix\url{https://doi.org/10.1073/pnas.2307151120}.

\bibitem{chen_competing_2019}
\bibinfo{author}{Chen, S.} \emph{et~al.}
\newblock \bibinfo{title}{Competing {Fractional} {Quantum} {Hall} and
  {Electron} {Solid} {Phases} in {Graphene}}.
\newblock \emph{\bibinfo{journal}{Physical Review Letters}}
  \textbf{\bibinfo{volume}{122}}, \bibinfo{pages}{026802}
  (\bibinfo{year}{2019}).
\newblock
  \urlprefix\url{https://link.aps.org/doi/10.1103/PhysRevLett.122.026802}.
\newblock \bibinfo{note}{Publisher: American Physical Society}.

\bibitem{das_sarma_perspectives_1996}
\bibinfo{author}{Das~Sarma, S.} \& \bibinfo{author}{Pinczuk, A.}
\newblock \emph{\bibinfo{title}{Perspectives in {Quantum} {Hall} {Effects}}}
  (\bibinfo{publisher}{John Wiley \& Sons, Ltd}, \bibinfo{year}{1996}),
  \bibinfo{edition}{1} edn.
\newblock
  \urlprefix\url{https://onlinelibrary.wiley.com/doi/10.1002/9783527617258}.
\newblock \bibinfo{note}{\_eprint:
  https://onlinelibrary.wiley.com/doi/pdf/10.1002/9783527617258}.

\bibitem{tesanovic_hall_1989}
\bibinfo{author}{Tešanović, Z.}, \bibinfo{author}{Axel, F.} \&
  \bibinfo{author}{Halperin, B.~I.}
\newblock \bibinfo{title}{‘‘{Hall} crystal’’ versus {Wigner} crystal}.
\newblock \emph{\bibinfo{journal}{Physical Review B}}
  \textbf{\bibinfo{volume}{39}}, \bibinfo{pages}{8525--8551}
  (\bibinfo{year}{1989}).
\newblock \urlprefix\url{https://link.aps.org/doi/10.1103/PhysRevB.39.8525}.

\bibitem{Note1}
\bibinfo{note}{The standard approach to remove the effect of mixing is to
  (anti)symmetrize the longitudinal resistivity (Hall resistance) as a function
  of magnetic field. Antisymmetrization could help reveal well-quantized
  plateaus in other Hall pairs. Given the time constraints of our measurement
  run at the National High Magnetic Field Lab, we were unable to acquire data
  for opposite field polarity.}

\bibitem{Wang2023_2}
\bibinfo{author}{Wang, X.} \& \bibinfo{author}{Vafek, O.}
\newblock \bibinfo{title}{Theory of correlated chern insulators in twisted
  bilayer graphene}.
\newblock \emph{\bibinfo{journal}{Phys. Rev. X}} \textbf{\bibinfo{volume}{14}},
  \bibinfo{pages}{021042} (\bibinfo{year}{2024}).
\newblock \urlprefix\url{https://link.aps.org/doi/10.1103/PhysRevX.14.021042}.

\bibitem{Vafek2023}
\bibinfo{author}{Vafek, O.} \& \bibinfo{author}{Kang, J.}
\newblock \bibinfo{title}{Continuum effective hamiltonian for graphene bilayers
  for an arbitrary smooth lattice deformation from microscopic theories}.
\newblock \emph{\bibinfo{journal}{Physical Review B}}
  \textbf{\bibinfo{volume}{107}} (\bibinfo{year}{2023}).
\newblock \urlprefix\url{http://dx.doi.org/10.1103/PhysRevB.107.075123}.

\bibitem{Kang2023}
\bibinfo{author}{Kang, J.} \& \bibinfo{author}{Vafek, O.}
\newblock \bibinfo{title}{Pseudomagnetic fields, particle-hole asymmetry, and
  microscopic effective continuum hamiltonians of twisted bilayer graphene}.
\newblock \emph{\bibinfo{journal}{Physical Review B}}
  \textbf{\bibinfo{volume}{107}} (\bibinfo{year}{2023}).
\newblock \urlprefix\url{http://dx.doi.org/10.1103/PhysRevB.107.075408}.

\bibitem{kang_analytical_2025}
\bibinfo{author}{Kang, J.} \& \bibinfo{author}{Vafek, O.}
\newblock \bibinfo{title}{Analytical solution for the relaxed atomic
  configuration of twisted bilayer graphene including heterostrain}
  (\bibinfo{year}{2025}).
\newblock \urlprefix\url{http://arxiv.org/abs/2502.15154}.
\newblock \bibinfo{note}{ArXiv:2502.15154 [cond-mat]}.

\bibitem{prada_dirac_2021}
\bibinfo{author}{Prada, M.}, \bibinfo{author}{Tiemann, L.},
  \bibinfo{author}{Sichau, J.} \& \bibinfo{author}{Blick, R.~H.}
\newblock \bibinfo{title}{Dirac imprints on the \$g\$-factor anisotropy in
  graphene}.
\newblock \emph{\bibinfo{journal}{Physical Review B}}
  \textbf{\bibinfo{volume}{104}}, \bibinfo{pages}{075401}
  (\bibinfo{year}{2021}).
\newblock \urlprefix\url{https://link.aps.org/doi/10.1103/PhysRevB.104.075401}.
\newblock \bibinfo{note}{Publisher: American Physical Society}.

\bibitem{Xie2020}
\bibinfo{author}{Xie, M.} \& \bibinfo{author}{MacDonald, A.~H.}
\newblock \bibinfo{title}{Nature of the correlated insulator states in twisted
  bilayer graphene}.
\newblock \emph{\bibinfo{journal}{Phys. Rev. Lett.}}
  \textbf{\bibinfo{volume}{124}}, \bibinfo{pages}{097601}
  (\bibinfo{year}{2020}).
\newblock
  \urlprefix\url{https://link.aps.org/doi/10.1103/PhysRevLett.124.097601}.

\bibitem{Park2021}
\bibinfo{author}{Park, J.~M.}, \bibinfo{author}{Cao, Y.},
  \bibinfo{author}{Watanabe, K.}, \bibinfo{author}{Taniguchi, T.} \&
  \bibinfo{author}{Jarillo-Herrero, P.}
\newblock \bibinfo{title}{Tunable strongly coupled superconductivity in
  magic-angle twisted trilayer graphene}.
\newblock \emph{\bibinfo{journal}{Nature}} \textbf{\bibinfo{volume}{590}},
  \bibinfo{pages}{249--255} (\bibinfo{year}{2021}).
\newblock \urlprefix\url{https://doi.org/10.1038%2Fs41586-021-03192-0}.

\bibitem{Wong2020}
\bibinfo{author}{Wong, D.} \emph{et~al.}
\newblock \bibinfo{title}{Cascade of electronic transitions in magic-angle
  twisted bilayer graphene}.
\newblock \emph{\bibinfo{journal}{Nature}} \textbf{\bibinfo{volume}{582}},
  \bibinfo{pages}{198–202} (\bibinfo{year}{2020}).
\newblock \urlprefix\url{http://dx.doi.org/10.1038/s41586-020-2339-0}.

\bibitem{Tomarken2019}
\bibinfo{author}{Tomarken, S.~L.} \emph{et~al.}
\newblock \bibinfo{title}{Electronic compressibility of magic-angle graphene
  superlattices}.
\newblock \emph{\bibinfo{journal}{Physical Review Letters}}
  \textbf{\bibinfo{volume}{123}} (\bibinfo{year}{2019}).
\newblock \urlprefix\url{http://dx.doi.org/10.1103/PhysRevLett.123.046601}.

\bibitem{Nuckolls2020}
\bibinfo{author}{Nuckolls, K.~P.} \emph{et~al.}
\newblock \bibinfo{title}{Strongly correlated chern insulators in magic-angle
  twisted bilayer graphene}.
\newblock \emph{\bibinfo{journal}{Nature}} \textbf{\bibinfo{volume}{588}},
  \bibinfo{pages}{610–615} (\bibinfo{year}{2020}).
\newblock \urlprefix\url{http://dx.doi.org/10.1038/s41586-020-3028-8}.

\bibitem{shingla_highly_2023}
\bibinfo{author}{Shingla, V.} \emph{et~al.}
\newblock \bibinfo{title}{A highly correlated topological bubble phase of
  composite fermions}.
\newblock \emph{\bibinfo{journal}{Nature Physics}}
  \textbf{\bibinfo{volume}{19}}, \bibinfo{pages}{689--693}
  (\bibinfo{year}{2023}).
\newblock \urlprefix\url{https://www.nature.com/articles/s41567-023-01939-2}.
\newblock \bibinfo{note}{Publisher: Nature Publishing Group}.

\bibitem{mcdonald_topological_1996}
\bibinfo{author}{McDonald, I.~A.} \& \bibinfo{author}{Haldane, F. D.~M.}
\newblock \bibinfo{title}{Topological phase transition in the
  {\textbackslash}ensuremath\{{\textbackslash}nu\}=2/3 quantum {Hall} effect}.
\newblock \emph{\bibinfo{journal}{Physical Review B}}
  \textbf{\bibinfo{volume}{53}}, \bibinfo{pages}{15845--15855}
  (\bibinfo{year}{1996}).
\newblock \urlprefix\url{https://link.aps.org/doi/10.1103/PhysRevB.53.15845}.
\newblock \bibinfo{note}{Publisher: American Physical Society}.

\bibitem{Wang2019}
\bibinfo{author}{Wang, P.} \emph{et~al.}
\newblock \bibinfo{title}{Piezo-driven sample rotation system with ultra-low
  electron temperature}.
\newblock \emph{\bibinfo{journal}{Review of Scientific Instruments}}
  \textbf{\bibinfo{volume}{90}} (\bibinfo{year}{2019}).
\newblock \urlprefix\url{https://doi.org/10.1063/1.5083994}.

\bibitem{Brown1964}
\bibinfo{author}{Brown, E.}
\newblock \bibinfo{title}{Bloch electrons in a uniform magnetic field}.
\newblock \emph{\bibinfo{journal}{Physical Review}}
  \textbf{\bibinfo{volume}{133}}, \bibinfo{pages}{A1038--A1044}
  (\bibinfo{year}{1964}).
\newblock \urlprefix\url{https://doi.org/10.1103/physrev.133.a1038}.

\bibitem{Zak1964}
\bibinfo{author}{Zak, J.}
\newblock \bibinfo{title}{Magnetic translation group}.
\newblock \emph{\bibinfo{journal}{Physical Review}}
  \textbf{\bibinfo{volume}{134}}, \bibinfo{pages}{A1602--A1606}
  (\bibinfo{year}{1964}).
\newblock \urlprefix\url{https://doi.org/10.1103/physrev.134.a1602}.

\bibitem{KrishnaKumar2017}
\bibinfo{author}{Kumar, R.~K.} \emph{et~al.}
\newblock \bibinfo{title}{High-temperature quantum oscillations caused by
  recurring bloch states in graphene superlattices}.
\newblock \emph{\bibinfo{journal}{Science}} \textbf{\bibinfo{volume}{357}},
  \bibinfo{pages}{181--184} (\bibinfo{year}{2017}).
\newblock \urlprefix\url{https://doi.org/10.1126/science.aal3357}.

\bibitem{KrishnaKumar2018}
\bibinfo{author}{Kumar, R.~K.} \emph{et~al.}
\newblock \bibinfo{title}{High-order fractal states in graphene superlattices}.
\newblock \emph{\bibinfo{journal}{Proceedings of the National Academy of
  Sciences}} \textbf{\bibinfo{volume}{115}}, \bibinfo{pages}{5135--5139}
  (\bibinfo{year}{2018}).
\newblock \urlprefix\url{https://doi.org/10.1073/pnas.1804572115}.

\bibitem{Barrier2020}
\bibinfo{author}{Barrier, J.} \emph{et~al.}
\newblock \bibinfo{title}{Long-range ballistic transport of brown-zak fermions
  in graphene superlattices}.
\newblock \emph{\bibinfo{journal}{Nature Communications}}
  \textbf{\bibinfo{volume}{11}} (\bibinfo{year}{2020}).
\newblock \urlprefix\url{https://doi.org/10.1038/s41467-020-19604-0}.

\bibitem{Kang2024}
\bibinfo{author}{Kang, J.} \& \bibinfo{author}{Vafek, O.}
\newblock \bibinfo{note}{In prep}.

\bibitem{Hejazi2019}
\bibinfo{author}{Hejazi, K.}, \bibinfo{author}{Liu, C.} \&
  \bibinfo{author}{Balents, L.}
\newblock \bibinfo{title}{Landau levels in twisted bilayer graphene and
  semiclassical orbits}.
\newblock \emph{\bibinfo{journal}{Physical Review B}}
  \textbf{\bibinfo{volume}{100}} (\bibinfo{year}{2019}).
\newblock \urlprefix\url{http://dx.doi.org/10.1103/PhysRevB.100.035115}.

\bibitem{Ledwith2020}
\bibinfo{author}{Ledwith, P.~J.}, \bibinfo{author}{Tarnopolsky, G.},
  \bibinfo{author}{Khalaf, E.} \& \bibinfo{author}{Vishwanath, A.}
\newblock \bibinfo{title}{Fractional chern insulator states in twisted bilayer
  graphene: An analytical approach}.
\newblock \emph{\bibinfo{journal}{Physical Review Research}}
  \textbf{\bibinfo{volume}{2}} (\bibinfo{year}{2020}).
\newblock \urlprefix\url{http://dx.doi.org/10.1103/PhysRevResearch.2.023237}.

\bibitem{Vakili2005}
\bibinfo{author}{Vakili, K.} \emph{et~al.}
\newblock \bibinfo{title}{Spin-dependent resistivity at transitions between
  integer quantum hall states}.
\newblock \emph{\bibinfo{journal}{Physical Review Letters}}
  \textbf{\bibinfo{volume}{94}} (\bibinfo{year}{2005}).
\newblock \urlprefix\url{http://dx.doi.org/10.1103/PhysRevLett.94.176402}.

\bibitem{Maryenko2015}
\bibinfo{author}{Maryenko, D.} \emph{et~al.}
\newblock \bibinfo{title}{Spin-selective electron quantum transport in
  nonmagnetic mgzno heterostructures}.
\newblock \emph{\bibinfo{journal}{Physical Review Letters}}
  \textbf{\bibinfo{volume}{115}} (\bibinfo{year}{2015}).
\newblock \urlprefix\url{http://dx.doi.org/10.1103/PhysRevLett.115.197601}.

\bibitem{shihSpinselectiveMagnetoconductivityWSe22025}
\bibinfo{author}{Shih, E.-M.} \emph{et~al.}
\newblock \bibinfo{title}{Spin-selective magneto-conductivity in {{WSe2}}}.
\newblock \emph{\bibinfo{journal}{Nature Physics}} \bibinfo{pages}{1--6}
  (\bibinfo{year}{2025}).
\newblock \urlprefix\url{https://doi.org/10.1038/s41567-025-02918-5}.

\end{thebibliography}

\renewcommand\thefigure{S\arabic{figure}}
\setcounter{figure}{0}
\renewcommand\thetable{S\arabic{table}}
\setcounter{table}{0}
\renewcommand\thesubsection{S\arabic{subsection}}
\setcounter{subsection}{0}
\setcounter{section}{0}  
\section*{Supplemental material}

\subsection{Transport measurements} \label{sec:noise}

We measured the sample in the SCM-32T system at the National High Magnetic Field Laboratory (MagLab). Although the hybrid magnet had a nominal maximum field of 32~T, as early users of the system we were restricted to a maximum field of 28~T. We used the top-loading dil fridge insert with a base temperature of 50~mK at the mixing chamber plate. The sample probe had 16 DC measurement wires, so we were not able to measure every contact pair in our device. The probe did not feature cold low-pass filters. We included low-pass filters (borrowed from another user and described in Ref.~\cite{Wang2019}) at the breakout box, however there was still RF pickup between the breakout box and the sample.

In addition to RF noise-induced blurring, our measurements also suffered from low-frequency noise within the measurement instruments themselves. This noise appeared in the current measurement for Fig.~\ref{fig:fci} (Fig.~\ref{fig:curr_noise}), but not in the associated voltage measurements for all densities/fields, meaning that the noise was not in the current itself. Thus, for Fig.~\ref{fig:fci}, we do not use the current measurement, but instead divide by a constant 1~nA.

We sourced 1~nA of current at near 5~Hz using an SR860, sourcing one volt across a 1~G$\Omega$ resistor. We used SR550 and 560 voltage preamplifiers and SR830 and SR860 lock-in amplifiers to measure voltages. We used an Ithaco 1211 current preamplifier and an SR860 to measure current.

We took the first set of measurements on many contact pairs at once (Figs.~\ref{fig:all_xx} and \ref{fig:all_xy}). The Hall measurements in Fig.~\ref{fig:fci} were taken in this set. We then found that only measuring one contact pair at a time greatly reduced noise. We therefore took the measurement in Fig.~\ref{fig:megafan}, consisting only of a single longitudinal resistivity measurement, separately.

\begin{figure*}
\centering
\includegraphics[width=6in]{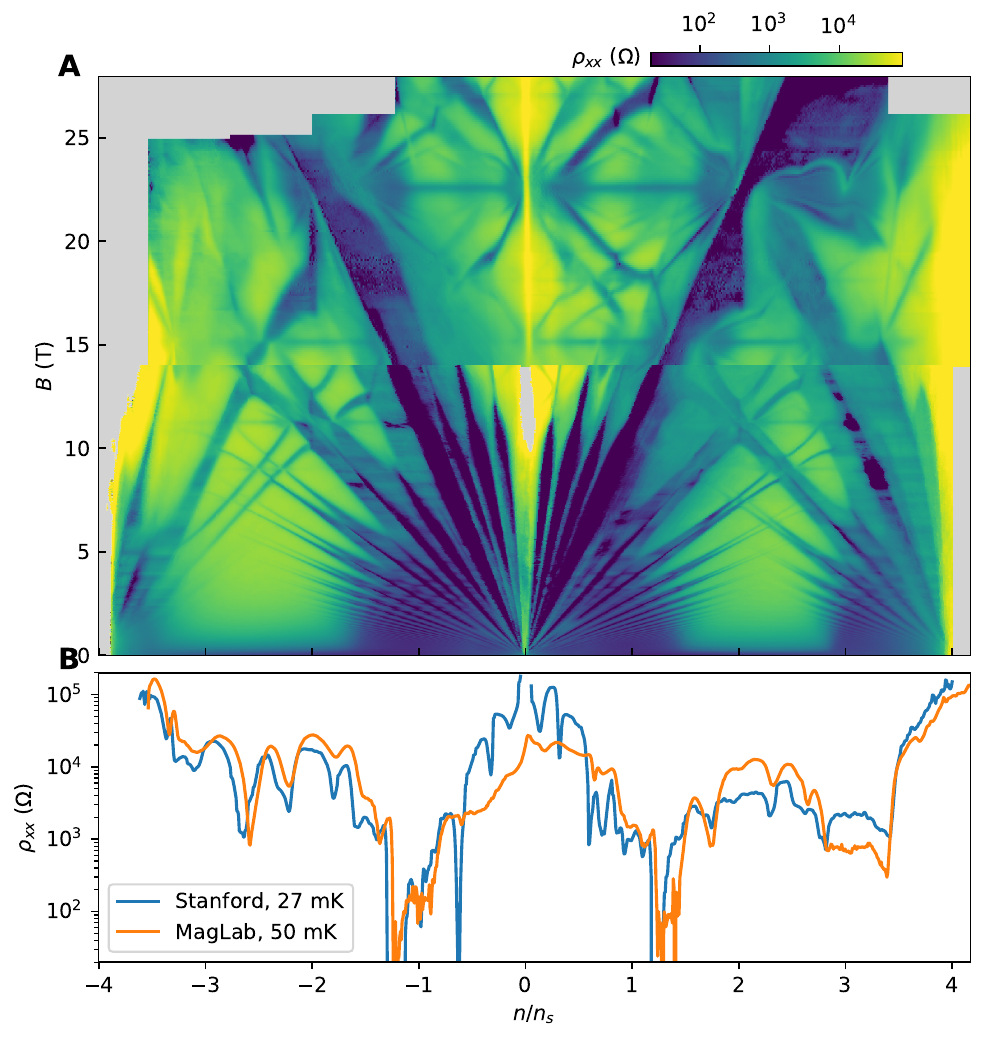}
\caption{\textbf{Full Landau fan including data taken at lower field.} Contact pair 7 - 8. (\textbf{A}) Data from 0 to 14~T are taken in our system at Stanford, previously published in Ref.~\cite{Finney2022}. Data above 14~T are the same as in Fig.~\ref{fig:megafan}, taken at MagLab. (\textbf{B}) Line cuts from both systems at 14~T.}
\label{fig:tlcomp}
\end{figure*}

We show a comparison between the data taken at MagLab with data taken in our own system at Stanford in Fig.~\ref{fig:tlcomp}. The system at Stanford has a lower base temperature and high quality filters at the mixing chamber stage. There are two stages of filtering: the wires are first passed through a cured mixture of epoxy and bronze powder to filter GHz frequencies, then low-pass RC filters mounted on sapphire plates filter MHz frequencies. As a result, the Landau fan appears much sharper. The resistivity near charge neutrality is nearly an order of magnitude higher in our measurements at Stanford compared to those at MagLab.

There is one notable difference between how the measurements were performed at Stanford versus MagLab. At Stanford, our contacts became highly resistive when the graphite back gate was held at ground, so we fixed the graphite back gate to 1.5~V for the measurements. At MagLab, we found that the contacts were an order of magnitude lower resistance than what we saw at Stanford, so we left the graphite back gate at ground. The difference in contact resistances was likely a result of a difference in effective temperature between the two setups.

\subsection{Generalized Diophantine and Definitions}
For concreteness, we explicitly follow the terminology used in \cite{tesanovic_hall_1989}, which we reproduce here. For a number of carriers per unit area $\bar{\rho}$, we write a generalized Diophantine equation as

\begin{equation}
    \bar{\rho} = \nu_s \rho_\phi + \eta_s A_0^{-1},
\end{equation}
where $\nu_s$ and $\eta_s$ are rational numbers, $\rho_\phi$ is the density of flux quanta per unit area, and $A_0$ is the unit cell area associated with the spatially modulated (crystalline) electron density. The Hall conductance is given by 

\begin{equation}
    \sigma_{xy} = - \nu_s e^2/h.
\end{equation}

A Wigner crystal ($\eta_s\ne0$, $\nu_s=0$) will have a spatially modulated electron density but vanishing Hall conductance.
An incompressible (fractional) quantum Hall liquid ($\eta_s=0$, $\nu_s\ne0$), and a Hall crystal ($\eta_s=0$, $\nu_s\ne0$) differ in that a quantum Hall liquid preserves translation symmetry whereas a Hall crystal spontaneously breaks translation symmetry with density fluctuations set by $\rho_\phi$ (an anomalous Hall crystal further spontaneously breaks time reversal symmetry). A Hall crystal differs from a Wigner crystal in having quantized Hall conductance: $\nu_s\ne0$.

A partial Hall crystal ($\nu_s\ne0$ and $\eta_s\ne0$) is an ``intermediate case'' combining the properties of both a Wigner crystal and a Hall crystal~\cite{tesanovic_hall_1989}. Therefore, it will have a non-zero Hall conductance and spatially modulated electron density. Given the Wigner crystal background, a partial Hall crystal will be compressible if the spatial modulation of the electron density is not strongly pinned to a disorder potential.

\subsection{Brown-Zak oscillations} \label{sec:bzo}

At rational values of $\Phi/\Phi_0=p/q$, the system can be described in terms of Bloch states in a $q$-times-enlarged magnetic Brillouin zone \cite{Brown1964,Zak1964}. In transport, this manifests as oscillations in conductivity at simple field fractions and constant density. These ``Brown-Zak'' oscillations have been observed in graphene/hBN superlattices\cite{KrishnaKumar2017,KrishnaKumar2018,Barrier2020}. The homogeneity and amount of disorder in the sample affects the maximum $q$ for which oscillations will be visible, because as $q$ increases, the size of the magnetic Brillouin zone increases relative to the carrier mean free path~\cite{KrishnaKumar2018}.

We observe Brown-Zak oscillations at $\Phi/\Phi_0=1/3$, $2/5$, $1/2$, and $3/5$ (Fig.~\ref{fig:megafan_sup} panel B, horizontal dotted lines). Of note, the Brown-Zak oscillations at $1/2$ flux appear only between $n/n_s=\pm2$. The amplitude of such oscillations depends on the carrier group velocity and therefore the miniband width~\cite{KrishnaKumar2018}, so absence of Brown-Zak oscillations in certain bands may indicate that those bands are flatter, with correspondingly greater influence of electron interactions. In general, we observe more and stronger Brown-Zak oscillations on the electron side of charge neutrality compared to the hole side, consistent with the familiar observation that in TBG the hole bands are flatter than electron bands.

We observe a similar set of Brown-Zak oscillations in contact pairs 16-17, 7-8, and 17-18 (Fig.~\ref{fig:all_xx}). Contact pair 6-7 has blurrier features overall, with Brown-Zak oscillations visible only at $\Phi/\Phi_0=1/2$ and $1/3$. Contact pair 13-14 does not show clear oscillations, likely because of both spatial inhomogeneity and stronger interactions brought by closer proximity to the magic angle.

\subsection{St\u{r}eda line fit procedure} \label{sec:fit}

The raw data for Figs.~\ref{fig:megafan}, \ref{fig:all_xx}, and \ref{fig:all_xy} were collected as a function of gate voltage and magnetic field. We transform the data into the space of $n/n_s$ vs $\Phi/\Phi_0$ by placing points along sharp lines with known $(s, t)$ by-eye. Brown-Zak oscillations additionally constrain $\Phi/\Phi_0$. We then use least squares to fit parameters $m_g$, $b_g$, and $B_0$, where $n/n_s=m_gV_g+b_g$ and $\Phi/\Phi_0=1$ at $B=B_0$. The resulting fit parameters are shown in Table~\ref{tab:fits} and schematically with our device geometry in Fig.~\ref{fig:schematic}. In this process we neglect the role of quantum capacitance, which depends on density of states and thus in turn on carrier density, making any global linear conversion between gate voltage and density imperfect.

After fitting the global conversion parameters to lines with known $(s, t)$, we were able to assign $(s, t)$ to the remaining straight-line features. Most of the lines could be easily assigned by eye, and the results for all contact pairs are shown in Fig.~\ref{fig:all_xx}. Not every relatively straight line in the space of density and flux along which $\rho_{xx}$ has a local minimum corresponds to a quantum Hall St\u{r}eda line, as already noted in Ref.~\cite{Finney2022}. In Figs.~\ref{fig:megafan_sup} and \ref{fig:all_xx} we attempted to label only St\u{r}eda lines.

\begin{table*}
\centering
\begin{tabular}{ c c c c c c }
& $m_g$ (V$^{-1}$) & $b_g$ & $B_0$ (T) & $A$ (nm$^2$) & $\theta$ (\textdegree) \\
\hline
13 - 14 & $0.843\pm0.005$ & $1.171\pm0.009$ & $41.2\pm0.4$ & $100.4\pm0.9$ & $1.309\pm0.006$ \\
6 - 16 & $0.723\pm0.003$ & $1.023\pm0.006$ & $48.0\pm0.3$ & $86.1\pm0.5$ & $1.414\pm0.004$ \\
6 - 7 & $0.716\pm0.002$ & $0.990\pm0.005$ & $48.2\pm0.2$ & $85.8\pm0.3$ & $1.416\pm0.002$ \\
16 - 17 & $0.774\pm0.002$ & $1.082\pm0.004$ & $45.1\pm0.1$ & $91.7\pm0.2$ & $1.369\pm0.002$ \\
7 - 17 & $0.772\pm0.002$ & $1.081\pm0.004$ & $45.2\pm0.2$ & $91.5\pm0.3$ & $1.372\pm0.003$ \\
7 - 8 & $0.771\pm0.002$ & $1.093\pm0.003$ & $45.1\pm0.1$ & $91.8\pm0.1$ & $1.369\pm0.001$ \\
7 - 8*& $0.770\pm0.001$ & $1.085\pm0.002$ & $45.2\pm0.1$ & $91.6\pm0.1$ & $1.371\pm0.001$ \\
17 - 18 & $0.769\pm0.002$ & $1.096\pm0.004$ & $45.6\pm0.1$ & $90.8\pm0.3$ & $1.377\pm0.002$ \\
8 - 18 & $0.734\pm0.007$ & $1.041\pm0.010$ & $47.6\pm0.7$ & $86.8\pm1.3$ & $1.408\pm0.011$ \\
\end{tabular}
\caption{\textbf{Voltage to density conversion parameters.} Fits to the data, primarily from Fig.~\ref{fig:all_xx}. The * marks a second measurement of contact pair 7 - 8 only, from Fig.~\ref{fig:megafan} -- we were able to decrease measurement noise by measuring only this one contact pair. It is important that $m_g$ and $B_0$ agree between the two measurements of contact pair 7 - 8, and they do. On the other hand, $b_g$ may vary between cooldowns as charges in the dielectrics or at interfaces migrate. In this table we report twist angles calculated assuming zero strain, despite the known presence of $\sim 0.2$\% uniaxial and biaxial heterostrain in the sample. At finite twist angle, moir\'e cell area depends on heterostrain only to second order in strain amount, but the influence of strain is magnified by the small twist angle~\cite{Wang2023}. Neglecting uniaxial heterostrain biases our extracted twist angle by 0.001 degrees, comparable to our reported fit uncertainties. The effect of 0.2\% biaxial heterostrain is substantially larger, biasing our extracted twist angle by 0.003 degrees, comparable and in some cases larger than our reported fit uncertainties. The unit cell area depends on biaxial homostrain to first order in strain amount, but without the small-angle magnification granted by heterostrain. Should biaxial homostrain be present in the sample at a level comparable to the heterostrain, neglecting it would bias our extracted twist angle by 0.001 degrees.}
\label{tab:fits}
\end{table*}

\subsection{Other contact pairs} \label{sec:otherpairs}

\begin{figure*}
\centering
\includegraphics[width=4in]{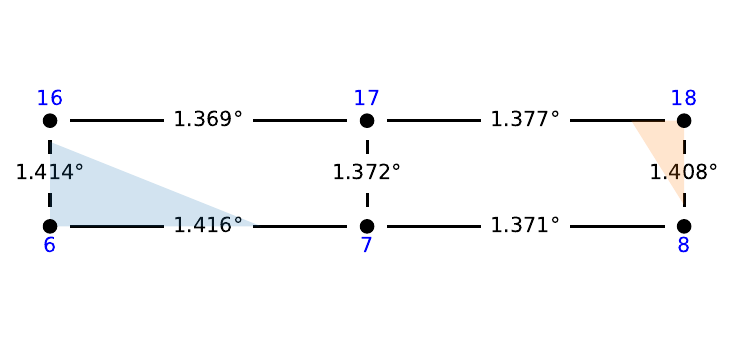}
\caption{\textbf{Twist angle variation.} Blue-labelled points are our contacts. The pitch between contacts is 3 $\mu$m. The channel is 1 $\mu$m wide. Vertical segments between them are Hall pairs, and horizontal segments are longitudinal pairs. The estimated $\theta$ values for each contact pair (Table~\ref{tab:fits}) are laid over the lines, with uncertainties elided for brevity. Shaded regions indicate our guess of regions with particularly inhomogeneous twist angles, based on nearby twist angles and the sharpness of features in the transport measurements.}
\label{fig:schematic}
\end{figure*}

Fig.~\ref{fig:schematic} shows a schematic of the device with measured twist angles overlaid. We can roughly identify where in the device twist angle is inhomogeneous, based on the sharpness of transport features and the spatial variation of measured twist angle. For instance, contact pairs 6~-~16 and 6~-~7 have high measured twist angle together with very blurry features in transport.

Hence, we draw the blue region. Contact pairs 17~-~18 is slightly blurry, and has a slightly higher twist angle than its neighbors, but contact pair 8~-~18 is very blurry and has a significantly higher twist angle, hence the orange region. These regions are only our best estimates. We do not have any local probes to measure them directly.

Significant differences between neighboring contact pairs has the unfortunate consequence that analyses that rely on comparing two measurements, typically $R_{xy}$ and $\rho_{xx}$, are not trustworthy.

\subsection{Single-particle bends, SBCI, Hofstadter ferrogmagnetism} \label{sec:sbci}

\begin{figure*}
\centering
\includegraphics[width=6in]{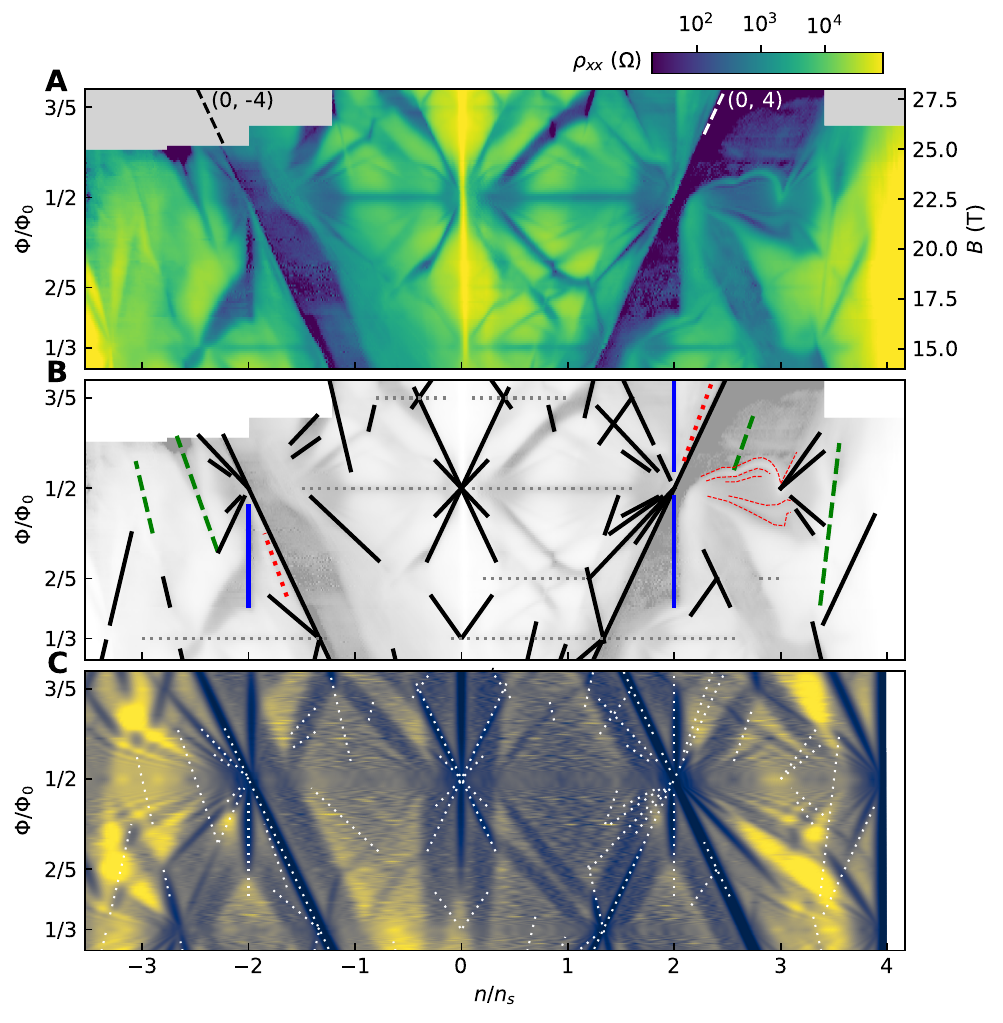}
\caption{\textbf{Longitudinal fan with schematic.} (\textbf{A}) Same data as in Fig.~\ref{fig:megafan}. (\textbf{B}) Black lines are ordinary integer quantum Hall. Blue vertical lines are potential QSH sites discussed in the main text. Green dashed lines are correlated Hofstadter ferromagnetic states (Sec.~\ref{sec:sbci}). Red dotted lines are symmetry-broken Chern insulating states (Sec.~\ref{sec:sbci}). Red dashed lines are the bending Landau levels discussed in Sec.~\ref{sec:main_llreset}. Gray dashed lines are Brown-Zak oscillations. (\textbf{C}) The solid black lines from B overlaid on the computation results as white dotted lines.}
\label{fig:megafan_sup}
\end{figure*}

As demonstrated in Ref.~\cite{Finney2022}, not all relatively straight lines in field/density with low resistance are St\u{r}eda gaps. Instead, relatively straight-line features of low resistivity can appear where one anisotropy-smeared butterfly spectrum is coincident with another. These false St\u{r}eda lines have a characteristic behavior at intersections with lines from the other doubled butterfly. Many of the features within our measurements follow the same pattern, and we attempt to only place lines in Fig.~\ref{fig:megafan_sup}B where we are confident that there is a gap.

\begin{figure*}
\centering
\includegraphics[width=6in]{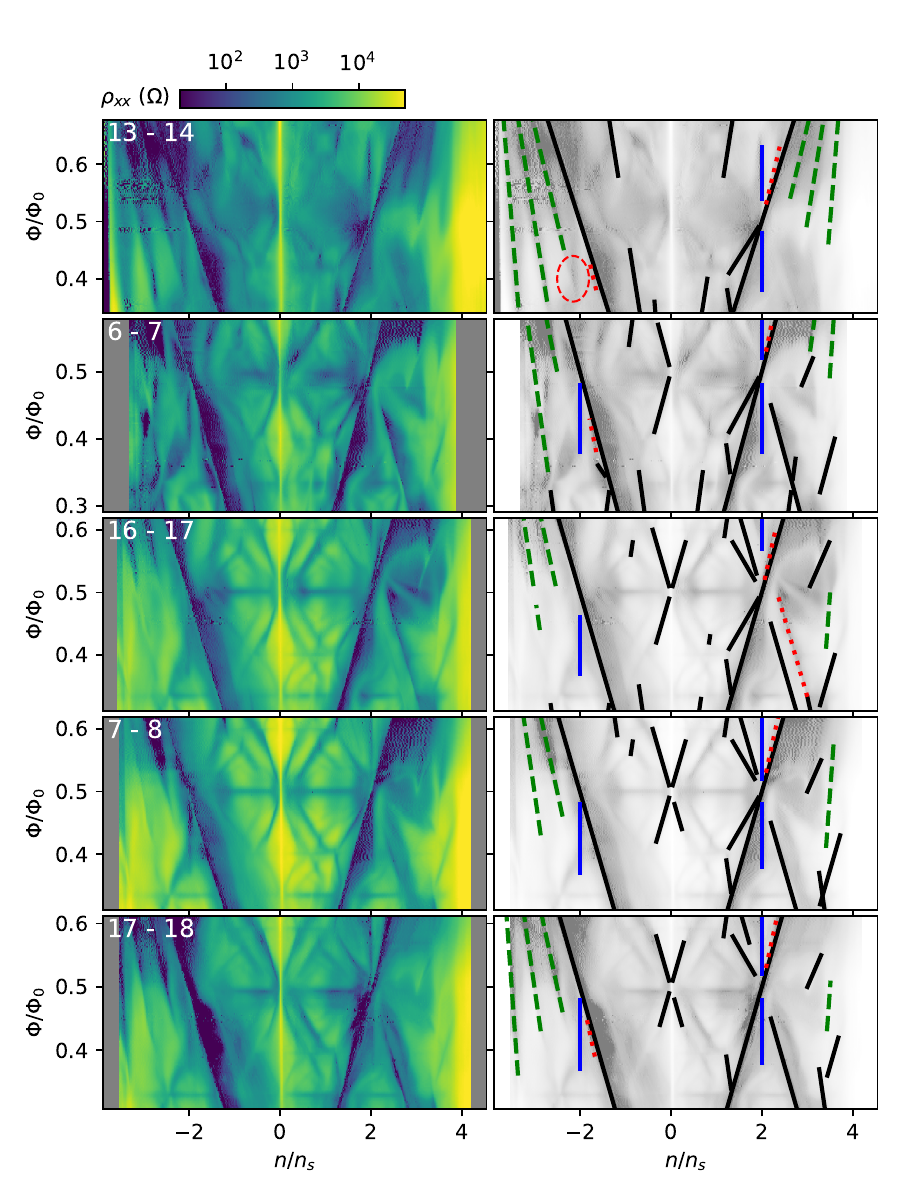}
\caption{\textbf{All longitudinal contact pairs.} Measured separately from Fig.~\ref{fig:megafan}, which was a remeasurement of contact pair 7-8. (\textbf{Left panels}) The longitudinal resistivity of the indicated contact pairs. (\textbf{Right panels}) Corresponding schematic with the same interpretation as in Fig.~\ref{fig:megafan_sup}B. The red circle for contact pair 13~-~14 indicates a feature that we could not unambiguously identify. It does not appear to be (-1, -3), which can be seen above it.}
\label{fig:all_xx}
\end{figure*}

\begin{figure*}
\centering
\includegraphics[width=6in]{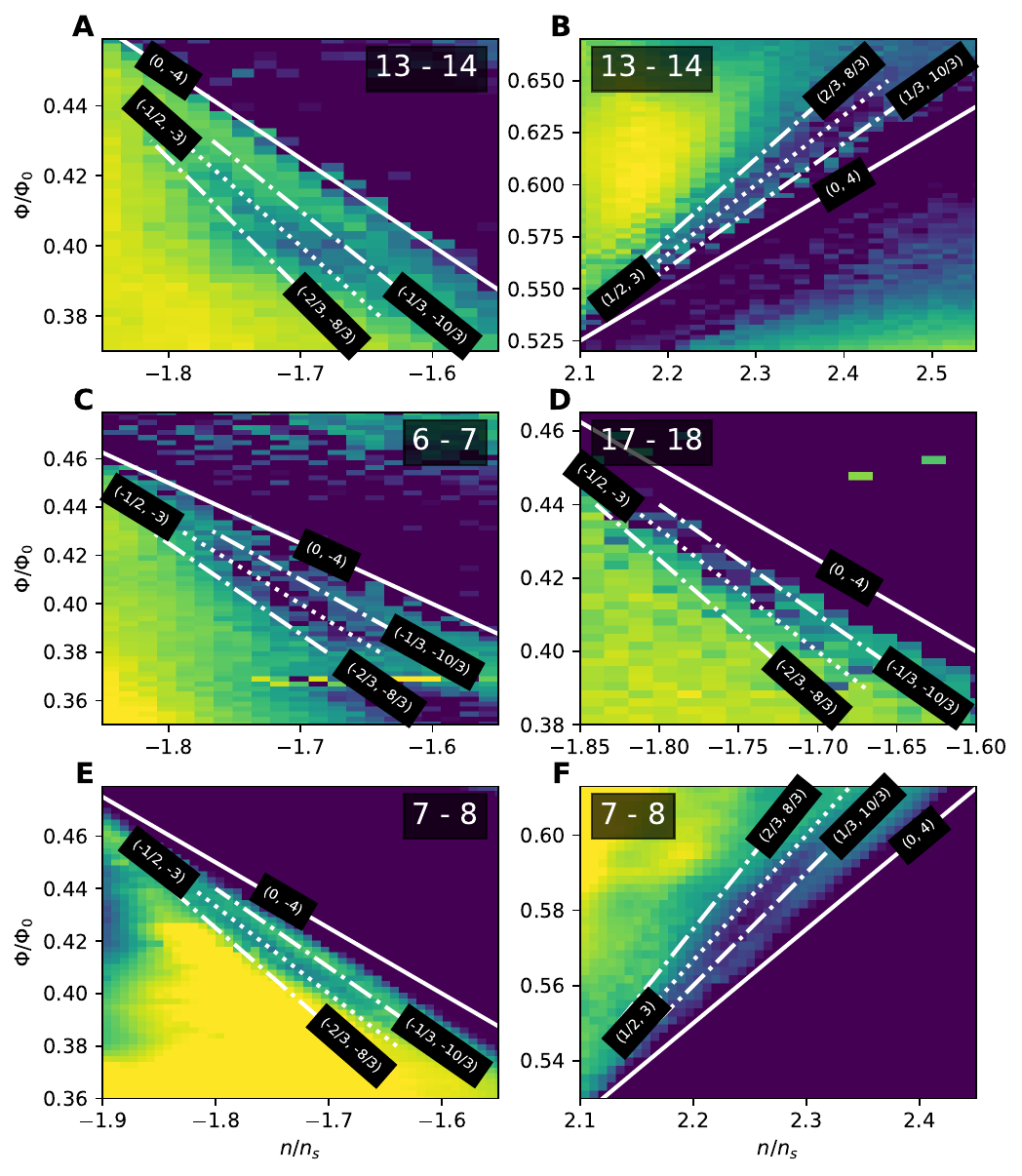}
\caption{\textbf{Fractional or symmetry-broken St\u{r}eda lines near half flux?} Panels \textbf{A}-\textbf{D} are the same data as in Fig.~\ref{fig:all_xx}. Panels \textbf{E}-\textbf{F} are the same data as in Fig.~\ref{fig:megafan}. Colormaps are adjusted for each panel to highlight the relevant line. Fits to $(\pm1/2, \pm3)$ (dotted lines) are better than fits to third fractions (dash dotted lines) in all cases except for panel~\textbf{F}.}
\label{fig:all_zoomies}
\end{figure*}

We observe a few St\u{r}eda lines with fractional $s=\pm 1/2$ and $t=\pm3$ near half flux (Fig.~\ref{fig:megafan_sup} red dotted lines in panel B and right panels of Fig.~\ref{fig:all_xx}). As they appear to emit from $n/n_s=\pm 2$ at half flux, we would describe them as $(\pm2, \pm3)_{1/2}$. They do not have a slope of $8/3$ (Fig.~\ref{fig:all_zoomies}). Such lines have been described as symmetry-broken Chern insulators in the literature \cite{Spanton2018}, and we observe at least one of them in all of the contact pairs that we measured. Curiously, in contact pair 16~-~17, a symmetry-broken St\u{r}eda line $(13/3, -4)=(3, -4)_{1/3}=(7/3, -4)_{1/2}$ runs between one third and one half flux.

We observe states with $s+t=\pm4$ projecting from integer $s$ (green dashed lines in Figs.~\ref{fig:megafan_sup} and \ref{fig:all_xx}). In all contact pairs, they tend to extend to lower field on the hole side compared to the electron side (Fig.~\ref{fig:all_xx}). In samples from the literature near the magic angle, they extend to zero or near zero field~\cite{Yankowitz2019,Tomarken2019,Park2021,Yu2022,Nuckolls2020,Pierce2021,Xie2021,Saito2021}. These lines are described theoretically as correlated Hofstadter ferromagnetic states in Ref.~\cite{Wang2023_2}, meaning they have polarized spin and valley degrees of freedom of interaction-modified bands.

\subsection{Detail of Landau level reset} \label{sec:resetzoom}

\begin{figure*}
\centering
\includegraphics[width=6in]{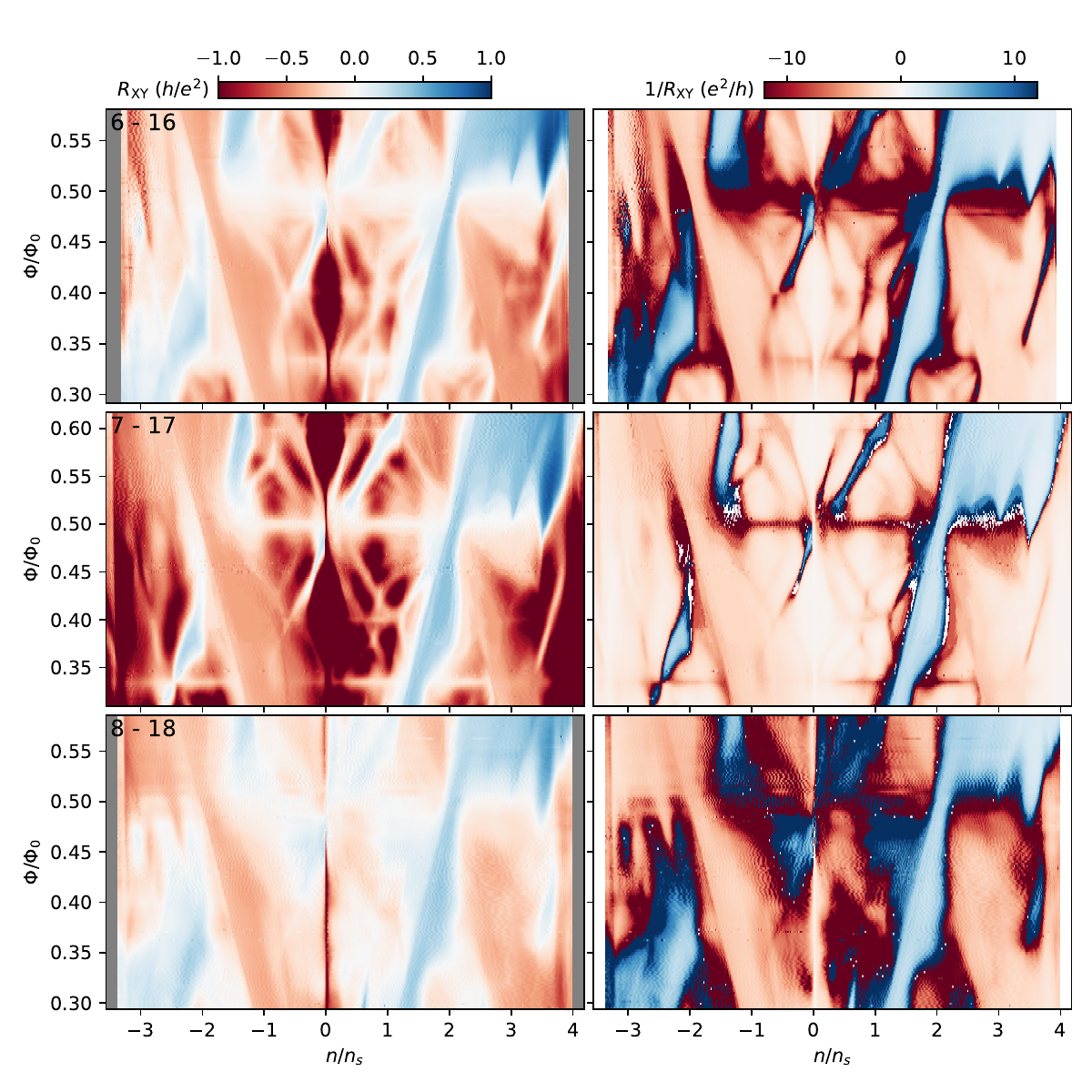}
\caption{\textbf{All Hall contact pairs.} (\textbf{Left panels}) $R_\mathrm{xy}$ for the indicated contact pairs. (\textbf{Right panels}) Corresponding $1/R_\mathrm{xy}$. Line cuts for each are shown in Fig.~\ref{fig:waterfallish}. Note that contact pairs 6~-~16 and 8~-~18 show significantly blurrier features compared to 7~-~17.}
\label{fig:all_xy}
\end{figure*}

\begin{figure*}
\centering
\includegraphics[width=6in]{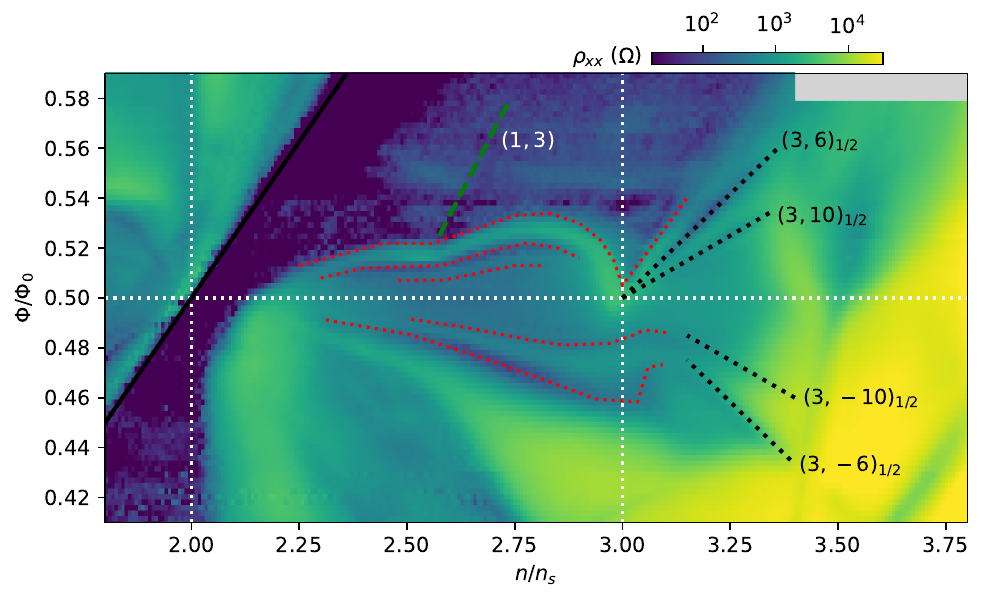}
\caption{\textbf{Detail of LL reset.} Data from Fig.~\ref{fig:megafan}. Other contact pairs shown in Figs.~\ref{fig:all_bendy} and \ref{fig:bendy_xy}. The green dashed line $(1, 3)$ is not prominently visible in the resistivity, but lines up with the kink in the red dotted line below it. The red dotted lines indicate minima in $\rho_\mathrm{XX}$ and likely correspond to Landau level gaps. The black dotted lines are minima that follow the indicated paths}
\label{fig:sharkbite}
\end{figure*}

\begin{figure*}
\centering
\includegraphics[width=4in]{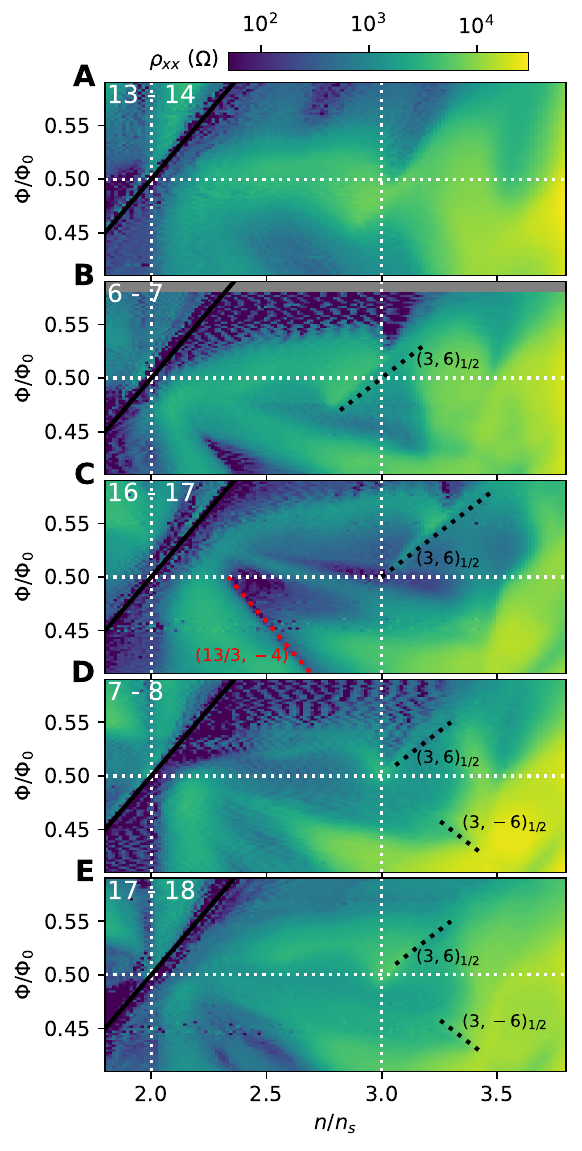}
\caption{\textbf{Detail of LL reset, continued.} All longitudinal contact pairs, labelled as in Fig.~\ref{fig:sharkbite}.}
\label{fig:all_bendy}
\end{figure*}

\begin{figure*}
\centering
\includegraphics[width=6in]{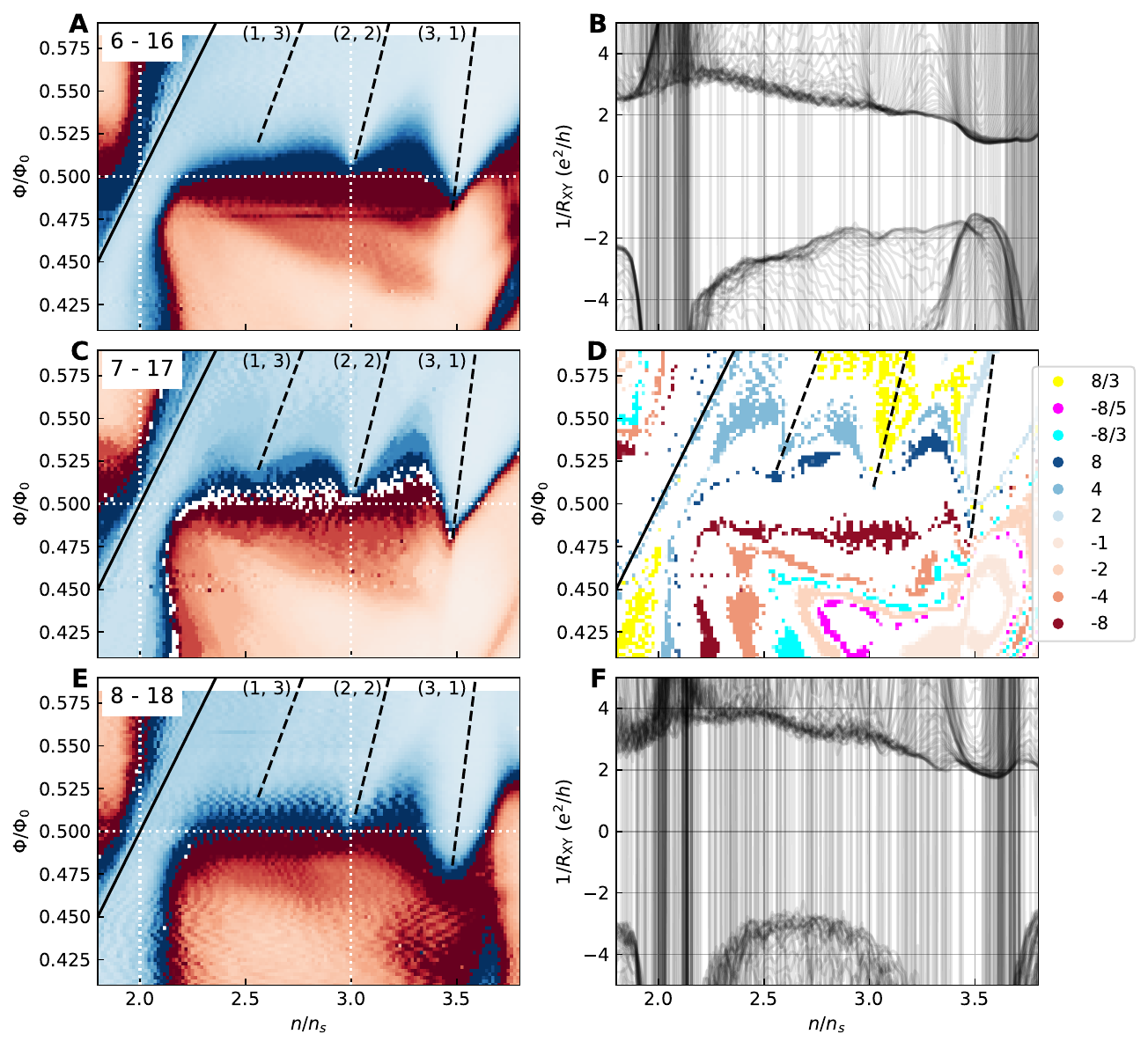}
\caption{\textbf{Detail of LL reset, continued.} All three Hall contact pairs show a change of sign near half flux (\textbf{ACE}). There appear to be features associated with the St\u{r}eda lines $(1, 3)$, $(2, 2)$, and $(3, 1)$, however the Hall conductivity never has the expected value for them (\textbf{BDF}). The sign change along the $(2, 2)$ line happens very close to half flux and $n/n_s=3$, however the sign change along $(3, 1)$ happens slightly below half flux in all cases.}
\label{fig:bendy_xy}
\end{figure*}

Figs.~\ref{fig:sharkbite}, \ref{fig:all_bendy}, and \ref{fig:bendy_xy} magnify the bending Landau minifan near $n/n_s=2$, $\Phi/\Phi_0=1/2$ from Figs.~\ref{fig:megafan}, \ref{fig:all_xx}, and \ref{fig:all_xy}.

In Fig.~\ref{fig:sharkbite}, red dotted lines mark bending, low-resistance states. The green dashed $(1, 3)$ line is not visible as a minimum in $\rho_{xx}$. Instead, we include it to highlight that there is a kink in the red dotted lines where it would intersect them. This line corresponds to an integer filling  of the band at half flux. After the reset at $n/n_s=3$ (half filling of the band), we observe St\u{r}eda lines $(3,\pm6)_{1/2}$ and $(3,\pm10)_{1/2}$.

The fourfold LL degeneracy is consistent with fully polarizing half of the bands at half flux. We note that above the topmost dotted red line, the transport behavior is the same as the spin-up regions described in the main text and Fig.~\ref{fig:spindept}. We suspect, therefore, that the band at half flux has fully spin-polarized at $n/n_s=3$.

To show that the Landau levels emanating from half flux are two-fold degenerate (per spin per valley), the simplest way to see this is to look at the directly at the Hofstadter spectrum. For example, when crossing from the (4, -4) to the (0, 4) gap slightly above or below half flux, we cross a single Hofstadter subband (in the absence of Zeeman splitting). This corresponds to a total change in the Chern number of 8, so a single Landau level must carry an extra two-fold degeneracy. 

To show this more concretely, we can start by considering a square lattice for simplicity. The effect of a magnetic field on the translation operators can be incorporated through a Peierls substitution
\begin{equation}
    T_x = e^{i(p_x +eA_x)x/\hbar},\ T_y = e^{i(p_y +eA_y)y/\hbar}
\end{equation}
where $A$ is the vector potential for some choice of gauge. For a given flux through the unit cell $\Phi/\Phi_0$, these operators will satisfy the commutation relation: 
\begin{equation}
    T_x T_y = T_y T_x e^{i 2\pi \Phi/\Phi_0}
\end{equation}
In the case where the flux per unit cell is a rational fraction $\Phi/\Phi_0=p/q$, for co-prime integers $p$ and $q$, this commutation relation can be used to show
\begin{equation}
    T_x^q T_y = T_y T_x^q;
\end{equation}
commutativity is restored over the enlarged magnetic unit cell. 

If we specifically consider the Landau gauge and consider a specific wavefunction $\psi_{n,\,p_y}$ that is an eigenstate of both the Hamiltonian and $T_y$. While this state is also an eigenstate of $T_x^q$, it is not an eigenstate of $T_x$. Therefore, there must exist some energetically degenerate state $T_x \psi_{n,\,p_y}$ with a different value of $p_y$. A state (or small cluster of states) that defines a low energy pocket of a band from which a sequence of Landau levels would form must be $q$-fold degenerate. In the case of half-flux, this means that a sequence of Landau levels has an extra two-fold degeneracy compared to a typical sequence emanating from zero field.

\subsection{Cluster filtering} \label{sec:clustering}
\begin{figure*}
\centering
\includegraphics[width=6in]{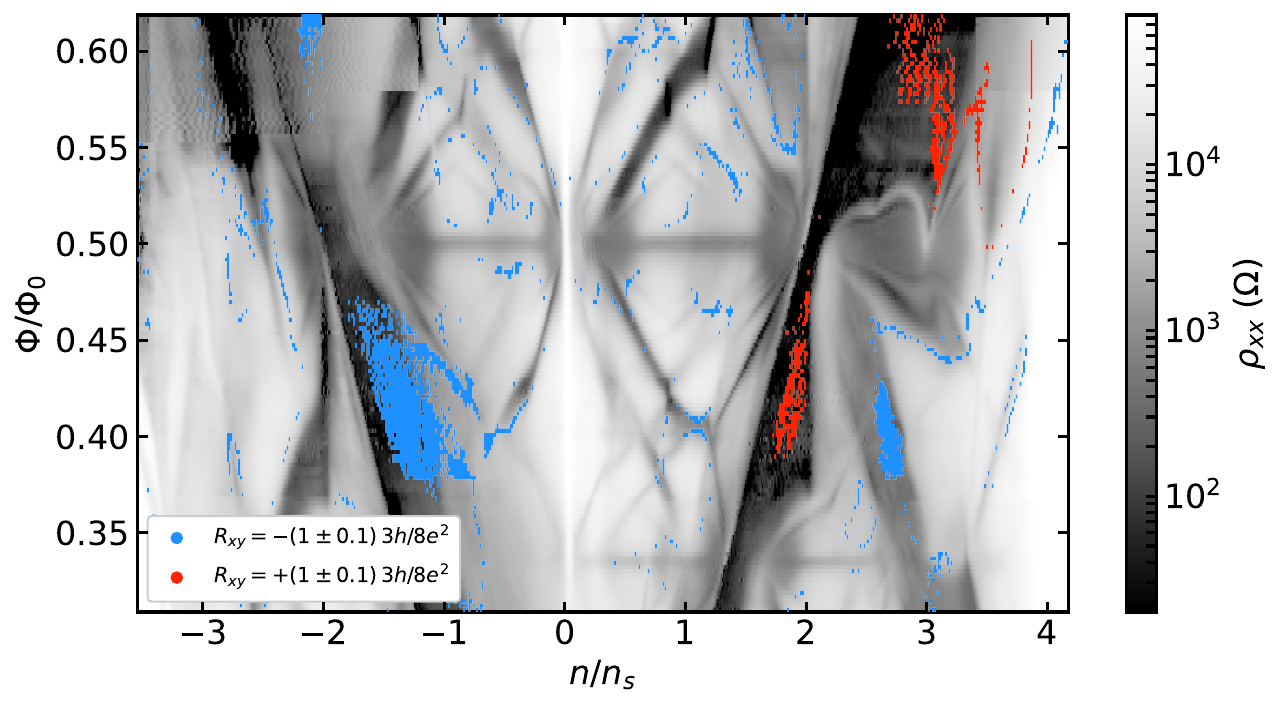}
\caption{\textbf{Raw Hall data.} 
Unfiltered version of Fig.~\ref{fig:fci}.
}
\label{fig:fci_raw}
\end{figure*}

The unfiltered data from Fig.~\ref{fig:fci} can be seen in Fig.~\ref{fig:fci_raw}. To filter the data, we begin by identifying points where the Hall resistance $R_{xy}$ falls within some window of our desired value ($\pm1\%$ of $\pm 3h/8e^2$ in this case). Binary dilation is applied to the windowed data such that nearest-neighbors within this masked array are now included. Each contiguous cluster of the dilated data set is then identified, and, if it does not contain above a threshold number of points, the original undilated points are omitted from the final filtered data set. Because of the dilation and generous threshold, this filtering primarily serves to remove isolated data points with no next-nearest-neighbor point within the window.

\subsection{How well quantized is the Hall resistance?} \label{sec:fci_quant}

\begin{figure*}
\centering
\includegraphics[width=0.5\textwidth]{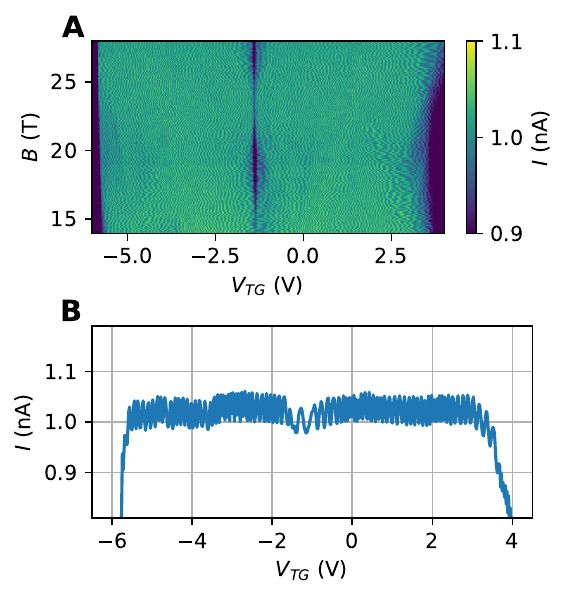}
\caption{\textbf{Current noise.} (\textbf{A}) Current vs gate and field. (\textbf{B}) Line cut at $B=14$~T. Note that the average current is a few percent above 1~nA.}
\label{fig:curr_noise}
\end{figure*}

\begin{figure*}
\centering
\includegraphics[width=6in]{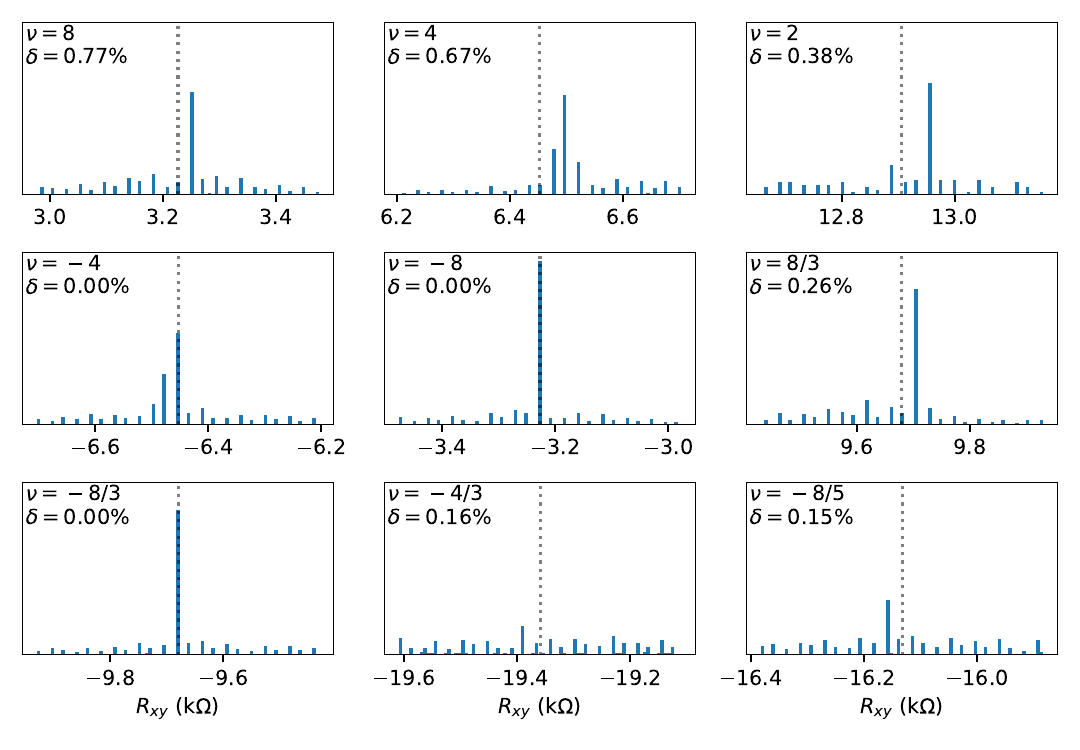}
\caption{\textbf{How well quantized is the Hall resistance?} Each panel shows a histogram (over the entire gate and field ranges) of $R_{xy}$ centered around the expected value of $h/\nu e^2$ (vertical dotted line) for $g=1.0785$ along with the computed $\delta$ value.}
\label{fig:fci_quant}
\end{figure*}

Fig.~\ref{fig:fci_quant} shows histograms of $R_{xy}=V_{xy}/(1~\mathrm{nA})/g$ near the indicated $\nu$ values. Quantum Hall plateaus show up as sharp peaks in this histogram. The free parameter $g$ represents inaccuracies in the gains of our amplifier and current source, which we did not measure carefully at MagLab. Then, we compute a metric of quantization
\begin{equation}
\delta=\frac{\mathrm{argmax}_R-R^*}{R^*}
\end{equation}
for each $\nu$ value. This metric is simple and does not consider, for instance, shoulders in the histogram. We show values of $\delta$ computed for $g=1.0785$. This number places $\nu=-8$, our most prominent quantum Hall plateau, directly on the expected value. It is instead possible to fine-tune $g$ to $1.0824$ so that all $\delta$ values are below 0.3\%. Either way, we are confident in stating that our fractions are quantized to better than 1\%. Note that 1.08 is a little high compared to our experience: typically the voltage pre-amps that we use at Stanford are only miscalibrated by a few percent. The 1.08 number also includes a factor from our current source, which we expect to be 1 to 2 percent based on the average of our noisy current (Fig.~\ref{fig:curr_noise}).

The simple $\delta$ metric does not consider the finite resolution of the voltage measurements: note that the histogram bins are typically spaced by a few tens of ohms. Assuming that it is $\sigma_{xy}$ that is quantized with
\begin{equation}
\sigma_{xy} = \frac{\nu e^2}{h} = \frac{R_{xy}}{R_{xy}^2 + \rho_{xx}^2},
\end{equation}
it is natural to ask how high can $\rho_{xx}$ be for $R_{xy}$ to still fall within the same histogram bin. For $\nu=\pm8/3$, the bins are roughly 20~$\Omega$ wide, yielding $|\rho_{xx}|\lesssim 300\;\Omega$. Indeed, in many of the places that $R_{xy}$ is precisely quantized to $\pm8/3$, $\rho_{xx}$ is of order a hundred ohms or less.

\subsection{Discussion of other fractions} \label{sec:other_fracs}
\begin{figure*}
\centering
\includegraphics[width=7in]{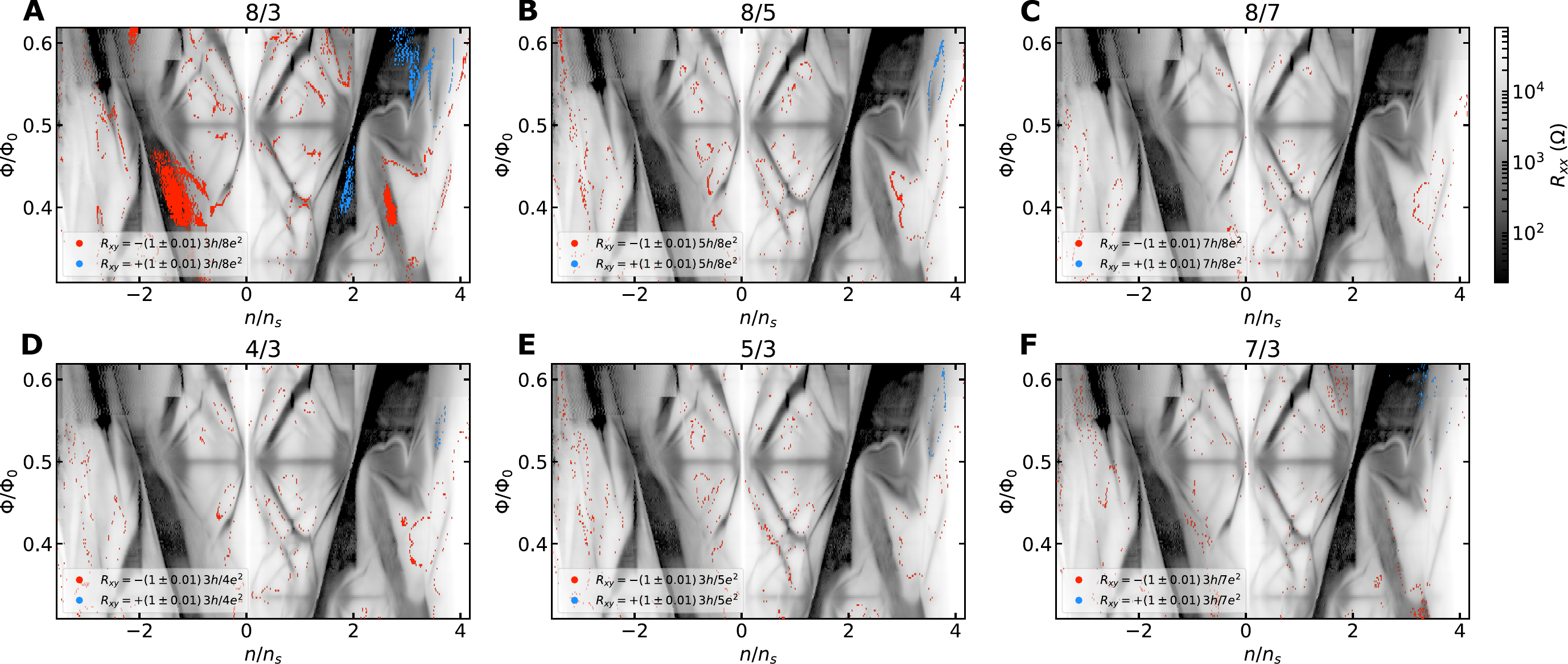}
\caption{\textbf{Highlighting values of Fractional Hall resistance} 
In a similar manner to Fig.~\ref{fig:fci}A of the main text, the longitudinal resistivity is shown in grey-scale. We then highlight regions where the Hall resistance falls within 1\% of the specified fractional value: \textbf{A} $\pm 8/3$ (reproduced from main text), \textbf{B} $\pm 8/5$, \textbf{C} $\pm 8/7$, \textbf{D} $4/3$, \textbf{E} $5/3$, and \textbf{F} 7/3.
Our clustering algorithm has been applied to remove spurious points (see Sec.~\ref{sec:clustering}). Note that the 1\% cutoff leads to different sizes of ranges of resistances for each fraction.
}
\label{fig:fci_all_fracs}
\end{figure*}

In a similar manner to Fig.~\ref{fig:fci}, in Fig.~\ref{fig:fci_all_fracs} we highlight where the Hall resistance $R_{xy}$ falls within a 1\% window about a specified fraction. We see that $\pm8/3$ is substantially more robust than the other displayed fractions. Most notably, $\-8/5$ and $-4/3$ exhibit the largest regions of quantization. Fractions not shown do not exhibit sizable regions that fall within a 1\% window. 

\subsection{Modeling a domain wall} \label{sec:domain_wall}

\begin{figure*}
\centering
\includegraphics[width=5in]{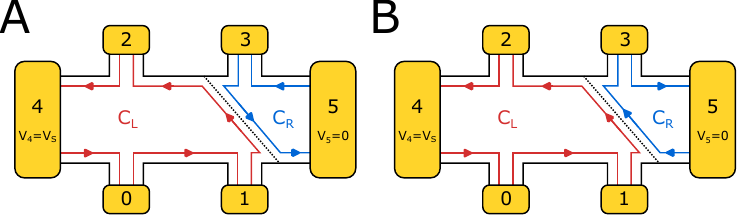}
\caption{\textbf{Hall bar with a domain wall.} Schematic diagram of a Hall bar with a domain wall between two Chern insulating regions with different Chern number $C_l$ and $C_r$ for the case of when (\textbf{A}) $\mathrm{sgn} (C_l)=\mathrm{sgn} (C_r)$ and (\textbf{B}) $\mathrm{sgn} (C_l)\ne\mathrm{sgn} (C_r)$.}
\label{fig:domain_wall}
\end{figure*}

To model the effects of a domain wall between two different Chern insulating regions with Chern numbers $C_l$ and $C_r$, we can use the Landauer-B\"uttiker formalism,
\begin{equation}
    I_i = \frac{e^2}{h} \sum_j (\bar{T}_{ji} V_j - \bar{T}_{ij} V_i),
\end{equation}
where the transmission coefficient $\bar{T}_{ij} = M_{ij}T_{ij}$ is the product of the number of modes $M_{ij}$ between contact j to contact i, and the transmittance $0 \le T_{ij} \le 1$. We will consider the case of perfect edge modes with some probability $t$ of a carrier initially starting in one edge mode ending in another. This will only have meaningful consequence for modes at the domain wall, where the carrier can end up in the adjacent domain. Let us first consider the case of a domain wall which passes in-between a Hall pair with $\mathrm{sgn} (C_l)=\mathrm{sgn} (C_r)$ (Fig.~\ref{fig:domain_wall}A). Assuming the source contact 4 is at voltage V and the drain contact 5 is at 0, for such a configuration, the full system of equations is

{\footnotesize
\[
\begin{pmatrix}
m_l & 0 & 0 & 0 & -m_l & 0 \\
-m_l & m_l & 0 & 0 & 0 & 0 \\
0 & -\bar{T}_{12} & m_l & -\bar{T}_{32} & 0 & 0 \\
0 & 0 & 0 & m_r & 0 & -m_r \\
0 & 0 & -m_l & 0 & m_l & 0 \\
0 & -\bar{T}_{15} & 0 & -\bar{T}_{35} & 0 & m_r
\end{pmatrix}
\times
\begin{pmatrix}
V \\ V \\ V_2 \\ 0 \\ V \\ 0
\end{pmatrix}
=
\begin{pmatrix}
0 \\ 0 \\ 0 \\ 0 \\ I \\ -I
\end{pmatrix},
\]
}
where $m_i = |C_i|$. Then 
\begin{align*}
    V_2 &= T_{12} V, \\
    I &= m_l V (1- T_{12}) = \bar{T}_{15} V.
\end{align*}

We can then solve for the expected resistance values: 
\begin{align*}
    R_{xx}^{\mathrm{bot}} &= \frac{V_0-V_1}{I} =0\\
    R_{xx}^{\mathrm{top}} &= \frac{V_2-V_3}{I}  = \frac{T_{12}}{m_l}\frac{1}{1-T_{12}}\\
    R_{yx}^{\mathrm{left}} &= \frac{V_0-V_2}{I} = \frac{1}{m_l}\\
    R_{yx}^{\mathrm{right}} &= \frac{V_1-V_3}{I} =  \frac{1}{m_l}\frac{1}{1-T_{12}}.
\end{align*}

We can similarly consider the case of a domain wall which passes in-between a Hall pair with $\mathrm{sgn} (C_l) \ne \mathrm{sgn} (C_r)$ (Fig.~\ref{fig:domain_wall}B). Again, assuming the source contact 4 is at voltage V and the drain contact 5 is at 0, for such a configuration, the full system of equations is

{\footnotesize
\[
\begin{pmatrix}
m_l & 0 & 0 & 0 & -m_l & 0 \\
-m_l & m_l & 0 & 0 & 0 & 0 \\
0 & -\bar{T}_{12} & m_l & 0 & 0 & -\bar{T}_{52} \\
0 & -\bar{T}_{13} & 0 & m_r & 0 & -\bar{T}_{53} \\
0 & 0 & -m_l & 0 & m_l & 0 \\
0 & 0 & 0 & -m_r & 0 & m_r
\end{pmatrix}
\times
\begin{pmatrix}
V \\ V \\ V_2 \\ V_3 \\ V \\ 0
\end{pmatrix}
=
\begin{pmatrix}
0 \\ 0 \\ 0 \\ 0 \\ I \\ -I
\end{pmatrix}.
\]}
For this case,  
\begin{align*}
    V_2 &= T_{12} V, \\
    V_3 &= T_{13} V, \\
    I &= m_l V (1- T_{12}).
\end{align*}
Solving for the expected resistance values: 
\begin{align*}
    R_{xx}^{\mathrm{bot}} &= \frac{V_0-V_1}{I} =0\\
    R_{xx}^{\mathrm{top}} &= \frac{V_2-V_3}{I}  = \frac{1}{m_l}\frac{T_{12}-T_{13}}{1-T_{12}}\\
    R_{yx}^{\mathrm{left}} &= \frac{V_0-V_2}{I} = \frac{1}{m_l}\\
    R_{yx}^{\mathrm{right}} &= \frac{V_1-V_3}{I} =  \frac{1}{m_l}\frac{1-T_{13}}{1-T_{12}}=\frac{1}{m_l}\frac{T_{12}}{1-T_{12}}.
\end{align*}

For either case, it would be possible to achieve a measured Hall resistance of 3/8 with the appropriate values of $m_l$ and $T_{12}$. However, such a scenario would yield markedly different values for $R_{xx}^{\mathrm{bot}}$ and $R_{xx}^{\mathrm{top}}$, which is inconsistent with our data. Therefore, we think such a domain wall in our sample is not the case. 

Following this same procedure, one can see that a domain wall that does not pass between a Hall pair cannot yield a fractionally quantized Hall conductance.

\subsection{Continuum Model and Quantum metric}
\label{supp:cont_model}

\label{sec:params}
In this subsection, we discuss the Hamiltonian of the continuum model in the presence of both heterostrain and out-of-plane magnetic field. Heterostrain is described by a $2 \times 2$ symmetric matrix $S_{\epsilon}$ which can always be diagonalized by an orthogonal matrix $R(\phi)$, 
\begin{align}
    S_{\epsilon} & = R_{\varphi}^T \begin{pmatrix} -\epsilon_\mathrm{uni} + \epsilon_\mathrm{bi} & 0\\
0 & \nu\epsilon_\mathrm{uni} + \epsilon_\mathrm{bi} \end{pmatrix}R_{\varphi}  \  . 
\end{align}
where  $R_{\varphi} = \begin{pmatrix} \cos\varphi & - \sin\varphi \\ \sin\varphi & \cos\varphi \end{pmatrix}$ is the two-dimensional rotational matrix with the angle of $\varphi$ that determines the two principal axes of the strain matrix $S_{\epsilon}$, and $R_{\varphi}^T$ is the transpose of the matrix $R_{\varphi}$, $\nu = 0.16$ is the Poisson ratio as used in Ref.~\cite{Wang2023}, and $\epsilon_\mathrm{uni}$ and $\epsilon_\mathrm{bi}$ are the uniaxial strain and biaxial strain respectively. In principle, the two layers may have different strain matrices $S_{\epsilon}^{(t)}$ and $S_{\epsilon}^{(b)}$, where the superscript $t$ and $b$ refers to top and bottom layers respectively. However, in this work, we consider only heterostrain with $S_{\epsilon} = S_{\epsilon}^{(t)} - S_{\epsilon}^{(b)}$ and neglect the homostrain $\frac12 (S_{\epsilon}^{(t)} + S_{\epsilon}^{(b)})$, since the shape of moire unit cell, the narrow band structure, the associated band properties, etc.~are more sensitive to heterostrain whose impact is estimated as the maximal value of $\epsilon/\theta$ with $\epsilon = \mathrm{max}(|\epsilon_{uni}|, |\epsilon_{bi}|)$ is the larger one between the uniaxial strain and the biaxial strain. On the other hand, the impact of homostrain is estimated as $\sim \epsilon$, much smaller than the heterostrain, and therefore can be safely neglected for small twist angles.

The impact of heterostrain on the non-interacting Hamiltonian have been discussed in Ref.~\cite{Wang2023,Vafek2023,Kang2023}, so we only briefly summarize the results in this manuscript. Following the model introduced in Ref.~\cite{Vafek2023,Kang2023}, we recalculated the lattice relaxation in the presence of the heterostrain by iteration method. We found that the change of the relaxation by the applied heterostrain is the order of $\epsilon/\theta$. To achieve the precision of $1$ meV in the dispersion of the narrow bands and include the particle-hole breaking terms, our continuum model contains all the terms up to the second order in the gradient expansion, with all the parameters numerically calculated from the microscopic tight binding model proposed in Ref.~\cite{Vafek2023,Kang2023}.  The Hamiltonian can be divided into two parts, the intralayer Hamiltonian $H_{\text{intra}}$ and the interlayer tunneling $H_{\text{inter}}$. The presence of heterostrain introduced several additional terms in $H_{\text{intra}}$. In addition to the coupling between the pseudo-scalar field induced by lattice distortion and the fermion density, $H_{\text{intra}}$ also includes the couplings between various strain-induced fields and the stress-energy of the Dirac fermion. $H_{\text{inter}}$ includes both the contact and gradient couplings. Since the tunneling constants, such as $w_0$, $w_1$, and $w_2$ in BM model, depends on both the microscopic interlayer hopping and the in-plane lattice distortion, and thus also depends on the applied strain. We have numerically calculated these coefficients with their values presented in Ref.~\cite{Kang2023,Kang2024}. 

As illustrated in Ref.~\cite{Wang2023}, the presence of heterostrain, in general, breaks the three-fold rotation $C_3$ and $C_{2x}$ symmetry, but still conserves $C_2 \mathcal{T}$. As a consequence, the two Dirac cones around the CNP are not gapped, but the Dirac points are shifted to other momenta in the moir\'e Brillouin zone, and the two Dirac points are not degenerate anymore, with the energy difference manipulated by heterostrain and observable by quantum oscillation experiments.

Crucially, the three degenerate van Hove points become non-degenerate as uniaxial heterostrain is introduced. In Ref.~\cite{Wang2023}, we measured the density of these non-degenerate van Hove points for both electrons and holes. Our model was electron/hole symmetric, so we could not fit to all six van Hove points simultaneously. Here, given we have now incorporated e-hole asymmetry, we can.

\begin{figure*}
\centering
\includegraphics[width=\textwidth]{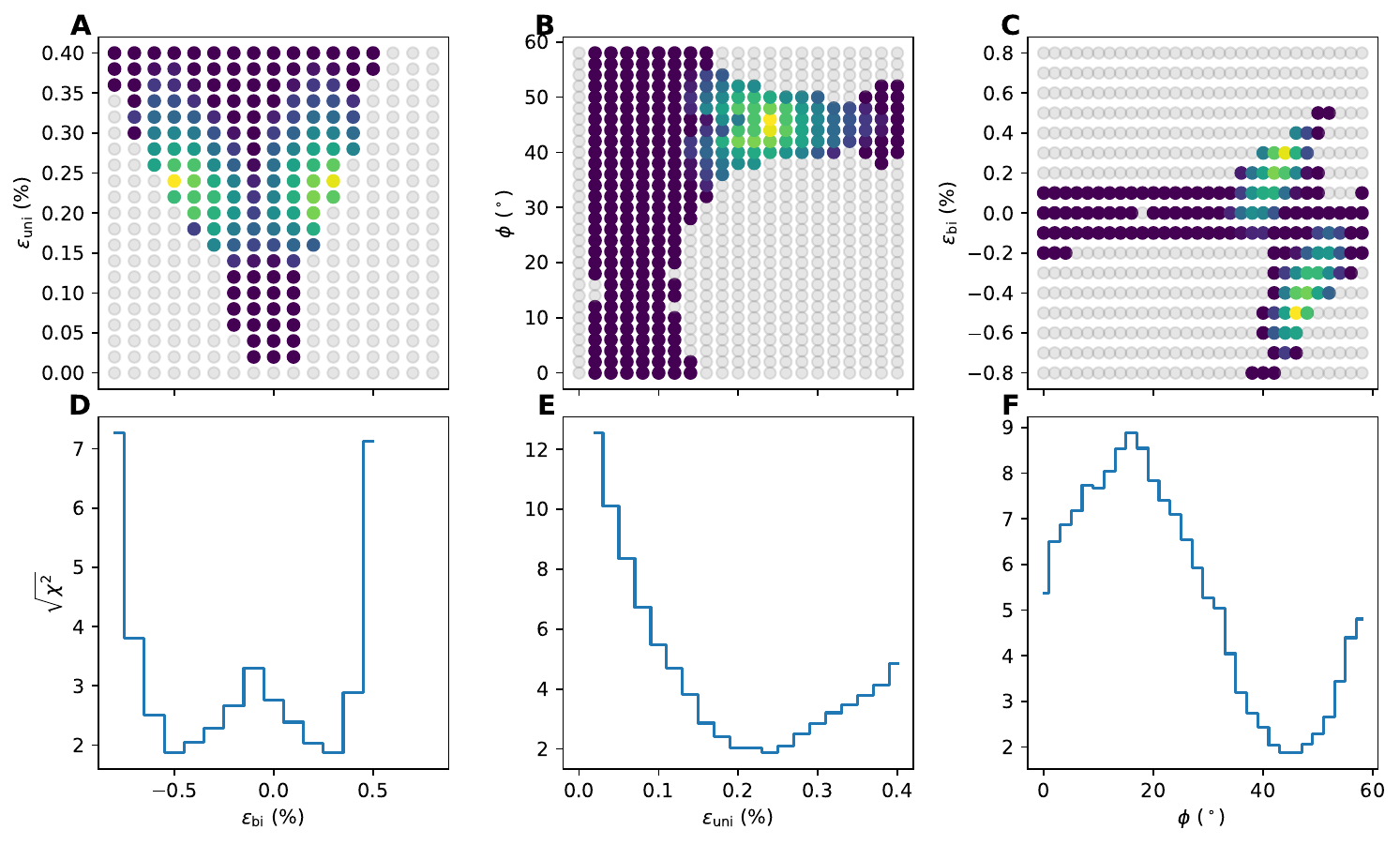}
\caption{\textbf{Best fit to low-field van Hove points.} (\textbf{A-C}) $\chi^2$ metric for the two indicated parameters (yellow is low, purple is high), using the best value of the not shown third dimension. (\textbf{D-F}) One dimensions $\chi^2$ metric, using the best value of both not shown dimensions.}
\label{fig:vh_fit}
\end{figure*}

Fig.~\ref{fig:vh_fit} shows the $\chi^2$ statistic for a wide variety of moire parameters $\epsilon_\mathrm{uni}$, $\epsilon_\mathrm{bi}$, and $\phi$. Given these three free parameters, we choose $\theta$ such that the moire unit cell area is fixed. In practice, fixing $\theta$ instead to 1.37\textdegree\ does not make a significant change to these results.

There are two minima in $\epsilon_\mathrm{bi}$, at -0.5\% and 0.3\%. These two minima correspond to the same Hamiltonian after accounting for symmetries, with their asymmetry about zero $\epsilon_\mathrm{bi}$ coming from the Poisson ratio (which converts uniaxial heterostrain into both uniaxial heterostrain and a small amount of biaxial heterostrain). Thus, there is a unique set of parameters yielding a best fit.

In Ref.~\cite{Wang2023}, we estimated an uncertainty of roughly $\delta\nu=\pm0.05$ on each van Hove point. Here, our best-fit result does satisfy this bound on each van Hove point. We hesitate to place uncertainties on our best-fit parameters, because we believe that they will be dominated by systematics. The model is not perfect, and it is not straightforward to account for model uncertainty. The uncertainties that we do report are based on the width of the minima in Fig.~\ref{fig:vh_fit}, and are underestimates.

\begin{figure*}[th]
\centering
\includegraphics[width=18cm]{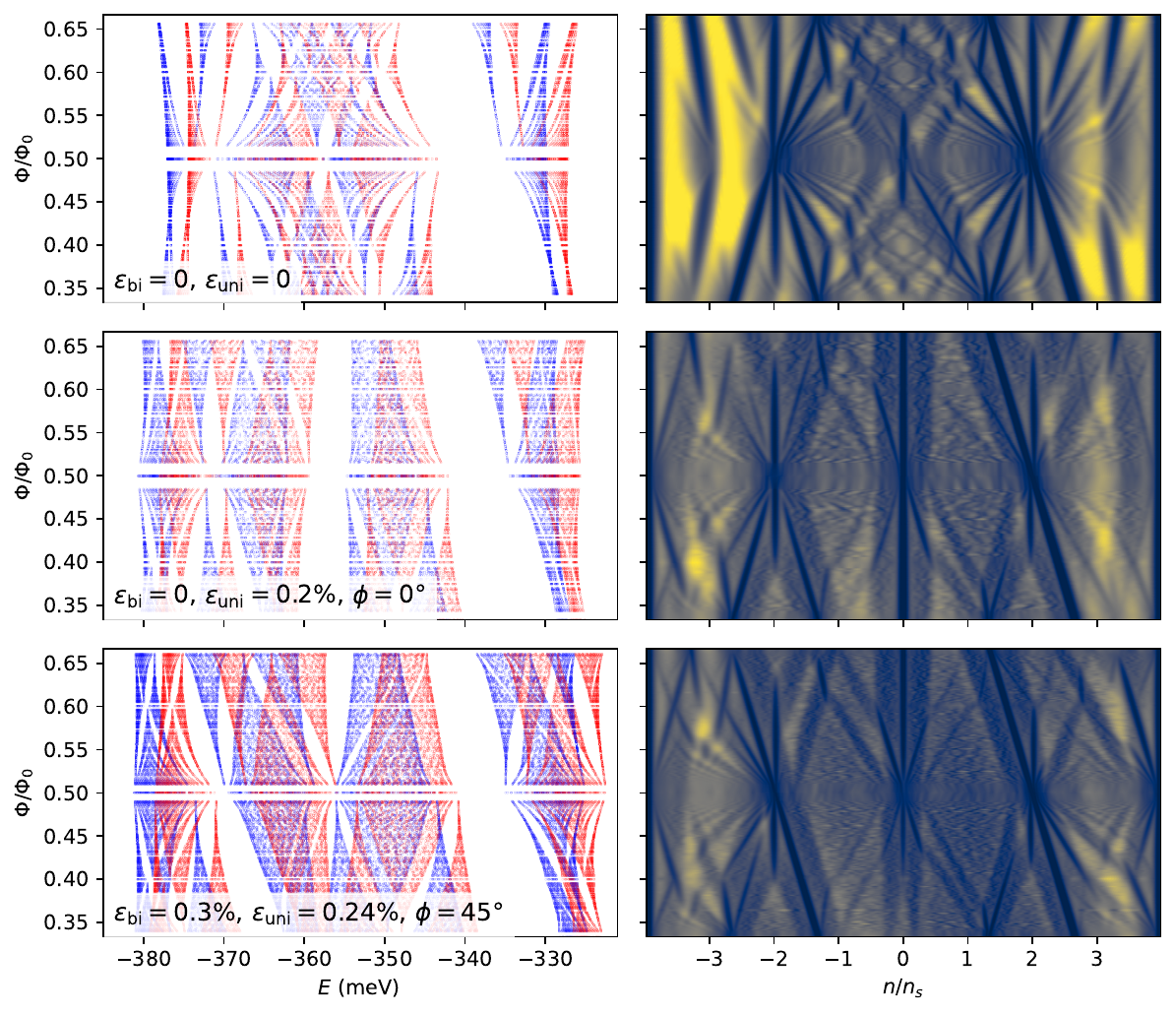}
\caption{\textbf{Computation comparison with different strain matrices.} (Left panels) Spectra for the indicated strain parameters. (Right panels) Associated Wannier plots.}
\label{fig:computation_compare}
\end{figure*}

The fits are only performed using low-field data. We find that at high field, the computed results look qualitatively similar for the amount of strain estimated in Ref.~\cite{Wang2023} and the amount fit here. See Fig.~\ref{fig:computation_compare}.

The Hamiltonian in the presence of a magnetic field can be obtained by substituting the operator $\hat{\fvec p} \longrightarrow \hat{\fvec p} + e \fvec A(\fvec x)$, with the Landau gauge $\fvec A = B(0, y)$. 

\begin{figure*}
\centering
\includegraphics[width=6in]{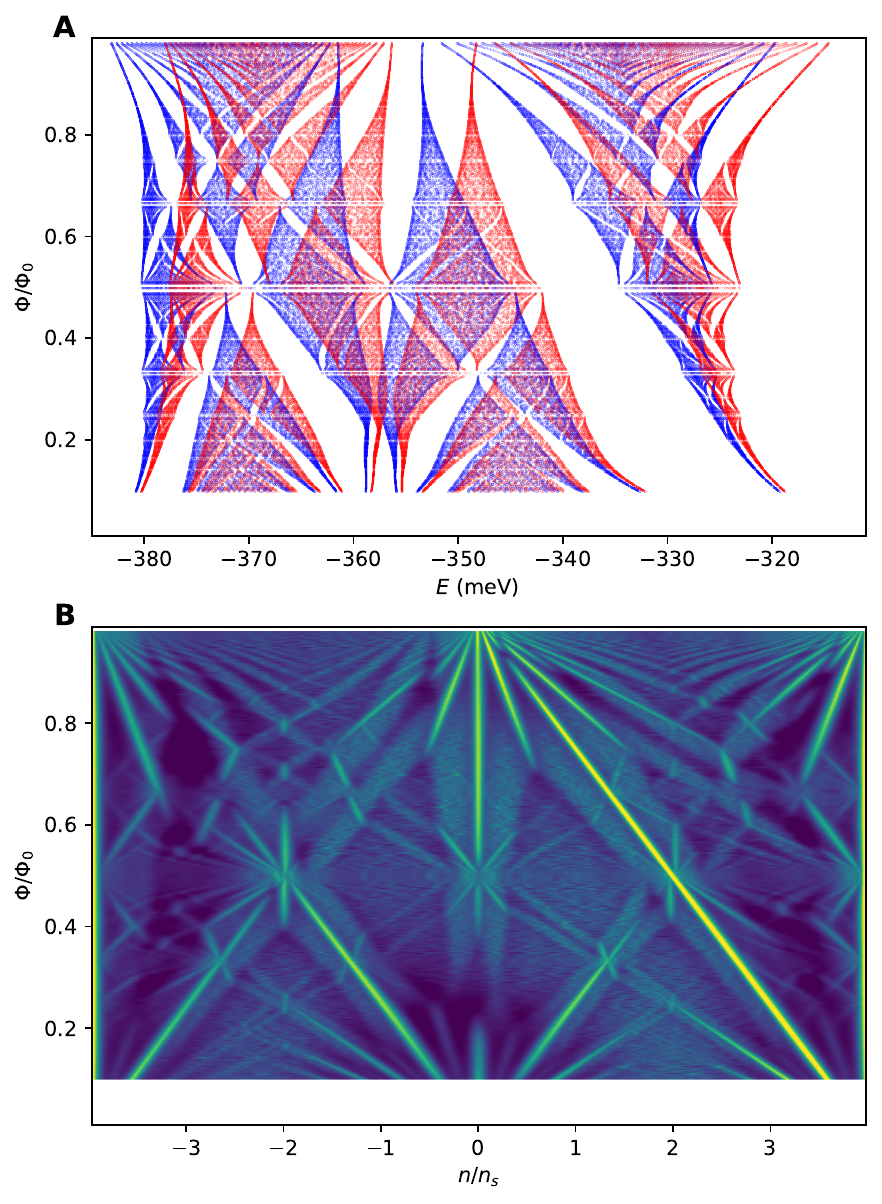}
\caption{\textbf{Hofstadter's butterfly of the strained BM model.} (\textbf{A}) Computed spectrum for $\theta=1.38^\circ$, uniaxial heterostrain $\epsilon=0.2$\% at 0\textdegree angle, and $g=2$. Red means spin down, blue means spin up. (\textbf{B}) Computed inverse density of states.}
\label{fig:computation_full}
\end{figure*}

For each field $p/q$, the computation produces a list of energies, eigenvalues of some large Hamiltonian. The number of energies produced is not the same from one field to another, growing with $q$. We then add a Zeeman term $E=\pm g\mu_BB/2$ for $g=2$ and $\mu_B$ is the Bohr magneton. Fig.~\ref{fig:computation_full}A shows the resulting spectrum.

To convert the list of energies at a given field $E_i$ into something that we can plot versus density, we follow Ref.~\cite{Hejazi2019} and write
\begin{equation*}
\rho(E)=\sum_i\frac{1}{\pi}\frac{\gamma}{(E-E_i)^2+\gamma^2},
\end{equation*}
\begin{equation*}
n(E)=\sum_i\frac{1}{\pi}\mathrm{arctan}\left(\frac{E-E_i}{\gamma}\right).
\end{equation*}
To normalize, divide each equation by the number of energies. The smoothing parameter $\gamma$ is not based on any temperature or disorder model. Fig.~\ref{fig:computation_full}B is then $1/\rho(E)$ plotted against $n(E)$ and field. It is refreshing to see that even with such a simple model, much of the behavior from both Ref.~\cite{Finney2022} and the present manuscript are captured.

At the end of this subsection, we briefly review the definition of Berry connection and quantum metric for the Bloch states. As usual, the overlap of two Bloch states are 
\begin{align}
    & \langle u(\fvec k ) | u(\fvec k + \delta \fvec k) \rangle =   1 + \delta k_{\mu} \langle u(\fvec k) | \partial_{\mu}  u(\fvec k) \rangle \nonumber \\
    & + \half \delta k_{\mu} \delta k_{\nu} \langle u(\fvec k) |  \partial_{\mu} \partial_{\nu} u(\fvec k) \rangle  + O(\delta k^3) \label{Eqn:BlochOverlap}
\end{align}
where $u(\fvec k)$ is the perodic part of the corresponding Bloch state, i.e.~$u(\fvec k) = e^{-i \fvec k \cdot \fvec r} \phi(\fvec k)$, where $\phi(\fvec k)$ is the Bloch state with the crystal momentum of $\fvec k$. Here, the overlap  $ \langle u(\fvec k + \delta \fvec k) | u(\fvec k) \rangle$ is expanded to the quadratic order of $\delta \fvec k$. It is obvious that the linear part of $\delta k_{\mu}$ is proportional to the Berry connection, defined as
\[  \mathcal{A}_{\mu} = i \langle u(\fvec k) | \partial_{\mu} u(\fvec k) \rangle  \]
and can be shown to be real. Additionally,
\begin{align}
    & \quad \langle u(\fvec k) |  \partial_{\mu} \partial_{\nu} u(\fvec k) \rangle  = -i \big( \partial_{\mu} \mathcal{A}_{\nu} + \partial_{\nu} \mathcal{A}_{\mu} \big) + \langle \partial_{\mu} \partial_{\nu} u (\fvec k) | u(\fvec k) \rangle \ .
\end{align}
Therefore, the imaginary part of $\langle u(\fvec k) |  \partial_{\mu} \partial_{\nu} u(\fvec k) \rangle$ is
\begin{align}
      - \half \big( \partial_{\mu} \mathcal{A}_{\nu} + \partial_{\nu} \mathcal{A}_{\mu} \big) 
\end{align}
However, it does not appear in the quadratic terms of the expansion  since it is an anti-symmetric tensor. We introduce the symmetric tensor $\gamma_{\mu\nu}$ for the real part
\begin{align}
   \gamma_{\mu\nu} = - \text{Re}(\langle u(\fvec k) |  \partial_{\mu} \partial_{\nu} u(\fvec k) \rangle)
\end{align}
Thus, Eq.~\ref{Eqn:BlochOverlap} can be rewritten as
\begin{align}
    & \quad \langle u(\fvec k ) | u(\fvec k + \delta \fvec k) \rangle  = 1 -i \mathcal{A}_{\mu} \delta k_{\mu} \nonumber \\
    & - \half \delta k_{\mu} \delta k_{\nu} \left(  \gamma_{\mu\nu} + \frac{i}2   \big( \partial_{\mu} \mathcal{A}_{\nu} + \partial_{\nu} \mathcal{A}_{\mu} \big)  \right) + O(\delta k^3)
\end{align}
However, the $\gamma$ tensor is $U(1)$ gauge dependent. To obtain a gauge-independent tensor, consider
\begin{align}
    & \quad \left|  \langle u(\fvec k ) | u(\fvec k + \delta \fvec k) \rangle  \right|^2 \nonumber \\
    & = 1 - \delta k_{\mu} \delta k_{\nu} \left(  \gamma_{\mu\nu} - \mathcal{A}_{\mu} \mathcal{A}_{\nu} \right) + O(\delta k^3)
\end{align}
and thus, the quantum metric tensor is defined as
\begin{align}
    & g_{\mu\nu}(\fvec k) = \gamma_{\mu\nu}(\fvec k) - \mathcal{A}_{\mu} \mathcal{A}_{\nu} \\
    \Longrightarrow & 1 -    \left|  \langle u(\fvec k ) | u(\fvec k + \delta \fvec k) \rangle  \right|^2 = \delta k_{\mu} \delta k_{\nu} g_{\mu\nu}(\fvec k)  + O(\delta k^3)
\end{align}
In the presence of magnetic fields,  the magnetic Bloch states might be written as the linear combination of Landau wave function instead of plane waves. In this case, the overlap between $u(\fvec k)$ and $u(\fvec k + \delta \fvec k)$ is calculated as
\begin{align}
    & \langle u(\fvec k ) | u(\fvec k + \delta \fvec k) \rangle \nonumber \\
    & = \frac1{S_{\text{uc}}} \int_{\text{muc}}\rmd^2 \fvec r\ \phi^*_{\fvec k}(\fvec r) e^{-i \delta \fvec k \cdot \fvec r} \phi_{\fvec k + \delta \fvec k}(\fvec r)
\end{align}
where $\int_{\text{muc}}\rmd^2 \fvec r$ integrates only over one magnetic unit cell, and $\phi_{\fvec k}$ is the magnetic Bloch state with the magnetic momentum of $\fvec k$.  All these formula will be discussed in detail in Ref.\cite{Kang2024}.

\begin{figure*}[th]
\centering
\includegraphics[width=18cm]{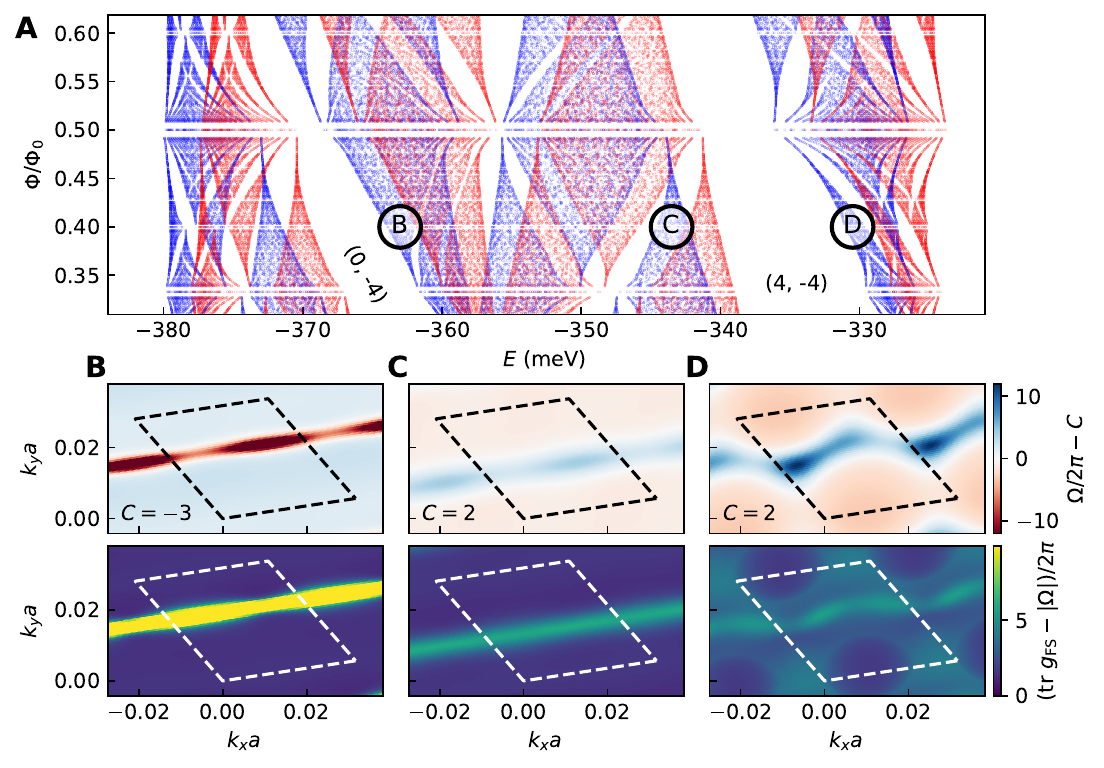}
\caption{\textbf{Computed Hofstadter spectrum and quantum geometric tensor.} (\textbf{A}) Computed energy levels for $\theta=1.35^\circ$ with $0.24\%$ uniaxial heterostrain at 45\textdegree\ and $0.3\%$ biaxial heterostrain ($q\leq72$), as described in the text. Spin up electrons are shown in blue, and spin down electrons are shown in red. The gaps at $(s, t)=(0, -4)$ and $(4, -4)$ are labeled. See Fig.~\ref{fig:computation_full} for the full range of fields. (\textbf{B}-\textbf{D}) Berry curvature (top panels) and trace condition (bottom panels) for the three labeled bands at 2/5 flux in panel A in the first moir\'e Brillouin zone. The black dashed lines correspond to the magnetic BZ. These three bands host the $-8/3$, $+8/3$, and $-8/3$ FCI states within the line cuts in Fig.~\ref{fig:fci}, respectively. The trace and Berry curvatures are dimensionless, having been scaled by the area of the magnetic BZ. The units are not scaled to a unit cell area of $\sqrt{3}\pi^2/2\approx8.5$}
\label{fig:qgt}
\end{figure*}

We show the Berry curvature and the trace of the quantum metric tensor for the bands featuring $\pm8/3$ FCIs at 2/5 flux in Fig.~\ref{fig:qgt}. We note that the ``trace condition''~\cite{Ledwith2020} is violated strongly in the case of one of these bands (panel B).

\subsection{Spin dependent transport detail} \label{sec:spindep}

There is a curious asymmetry in the longitudinal transport measurements. Observe the value of the resistivity on either side of $(0, \pm4)$ in Fig.~\ref{fig:megafan}. On the left side (lower density), the resistivity tends to be in the kilohms, and there are numerous St\u{r}eda lines immediately next to the $(0, \pm4)$ line (See Sec.~\ref{sec:spindep} for more detail). On the right side (higher density), the resistivity tends to be lower than a kiloohm, and in some cases it is below one Ohm. In these regions, there are no St\u{r}eda lines, and the resistivity is flat and featureless.

\begin{figure*}[th]
\centering
\includegraphics[width=8.5cm]{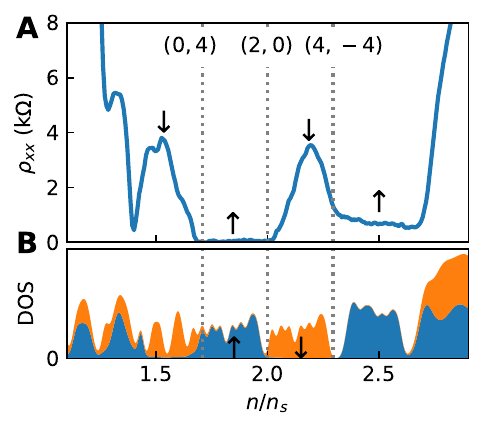}
\caption{\textbf{Possible spin-dependent transport.} (\textbf{A}) Line cut from Fig.~\ref{fig:megafan} at 19.2~T. Intersections with the $(0, 4)$, $(2, 0)$, and $(4, -4)$ Hofstadter gaps are indicated with vertical dotted lines. (\textbf{B}) Density of states at $p/q=19/45$ ($B\approx19.1$~T). Spin up and down states are indicated in blue and orange respectively, as well as with arrows in both panels.}
\label{fig:spindept}
\end{figure*}

As an example, a line cut at constant field is shown in Fig.~\ref{fig:spindept} panel A. To the left of $(0, 4)$, the resistivity is peaks around 4~kilohms. To the right, between $(0, 4)$ and $(2, 0)$, it is only a hundred Ohms or lower. In panel B we show that these two types of transport correspond to states in the model that are entirely spin-polarized. Where the states are spin-up, we see low resistivity, and where they are spin-down we see high resistivity. Where we see a mixture of both, the resistivity is even higher. This behavior is followed in our measurements not just near $(0, \pm4)$, but also in other regimes where the model reasonably matches the experiment. The major exception to this rule is near charge neutrality, where no substantial asymmetry is apparent.

Curiously, we only see FCI states in these low-resistivity (spin-up) regions. Circled $\alpha$, $\beta$, and $\gamma$ in Fig.~\ref{fig:computation} panels A and B indicate regions where we observe $\pm8/3$ FCI states. All of these regions are within blue Chern bands.

Spin-dependent resistivity has been observed in quantum Hall systems previously~\cite{Vakili2005,Maryenko2015,shihSpinselectiveMagnetoconductivityWSe22025}. With Zeeman splitting, the lower-energy, spin-up states are filled first. Spin-down electrons then see a different magnetic environment: there is a substantial background of spin-up electrons already present. In high-quality WSe$_2$, this leads to the minority-spin resistance being dramatically increased at low temperatures~\cite{shihSpinselectiveMagnetoconductivityWSe22025}. We observe the same behavior here, however we are not confident that the underlying explanation is the same. We may be able to reverse the direction of resistivity with back gate voltage. See Fig.~S3 panels C and D from Ref.~\cite{Finney2022}, where oscillations appear as a function of displacement field (y-axis). We wish to emphasize that we do not have positive evidence for the hypothesis of spin-dependent transport.

\begin{figure*}
\centering
\includegraphics[width=6in]{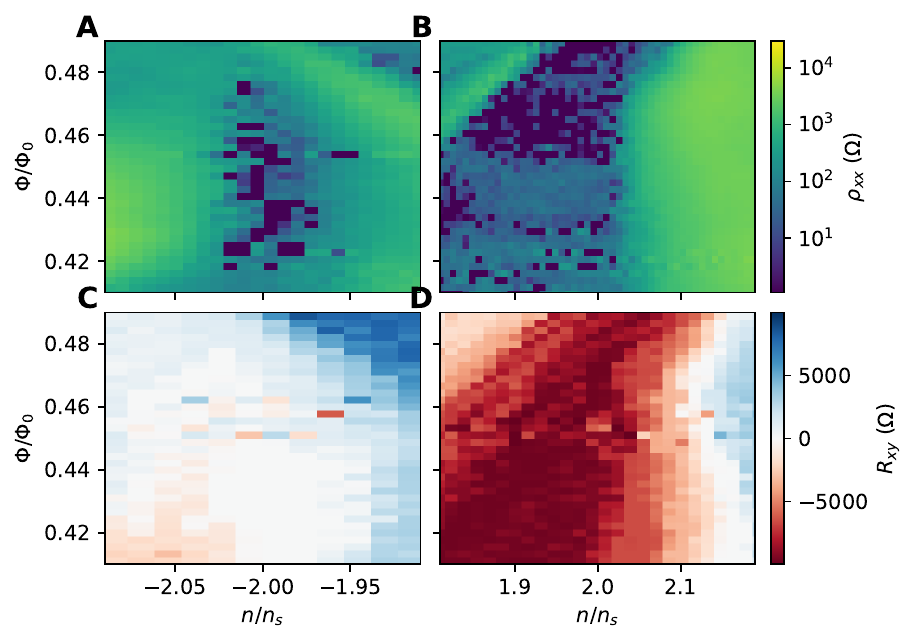}
\caption{\textbf{Fine detail of potential QSH.} (\textbf{A}-\textbf{B}) Zoom in on Fig.~\ref{fig:megafan} near the clear vertical-line features. (\textbf{C}-\textbf{D}) Same, for contact pair 7~-~17.}
\label{fig:qsh_megafan}
\end{figure*}

\begin{figure*}
\centering
\includegraphics[width=6in]{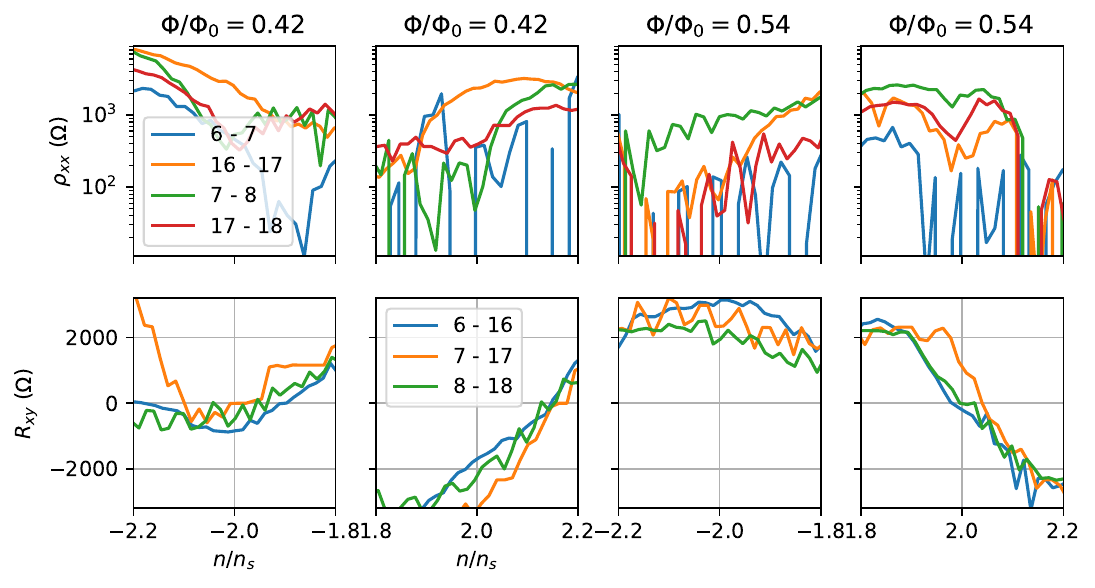}
\caption{\textbf{Line cuts near potential QSH.} Top row: line cuts of longitudinal measurements (Fig.~\ref{fig:all_xx}) at the indicated fields. Although there are some minima at $n/n_s=\pm2$, there is too much noise in most measurements to clearly claim a certain behavior. Bottom row: ditto, but for Hall measurements (Fig.~\ref{fig:all_xy}). Only contact pair 7~-~17 shows a clear plateau at zero resistance at $n/n_s=-2$ and $\Phi/\Phi_0=0.42$.}
\label{fig:qsh_cuts}
\end{figure*}

\begin{figure*}
\centering
\includegraphics[width=6in]{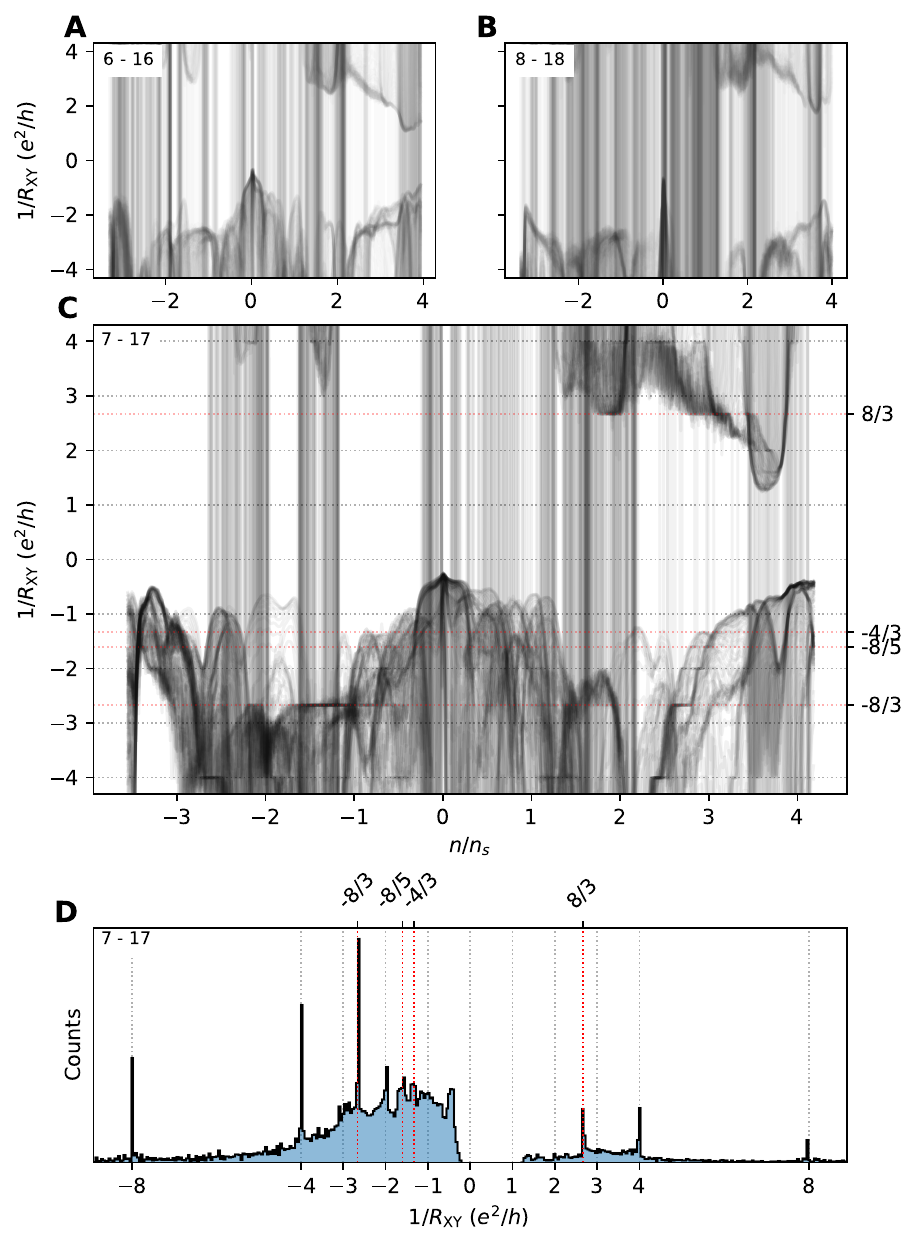}
\caption{\textbf{Hall plateaus.} All line cuts at constant field of a Landau fan for a given contact pair plotted over each other with transparency. Quantized Hall plateaus will appear as solid horizontal lines. Contact pairs indicated in the top left corner. (\textbf{A}, \textbf{B}) There are no quantized Hall plateaus for these contact pairs, not even integers. (\textbf{C}) Contact pair 7~-~17 shows quantized plateaus at integers $\pm8$ (not shown), $\pm4$, -2, -8/3, -8/5, and -4/3, the latter two only near $n/n_s=3$. (\textbf{D}) Histogram of all Hall conductances from panel B. Interestingly, there are more -8/3 measurements than any integer.}
\label{fig:waterfallish}
\end{figure*}

\begin{figure*}
\centering
\includegraphics[width=5in]{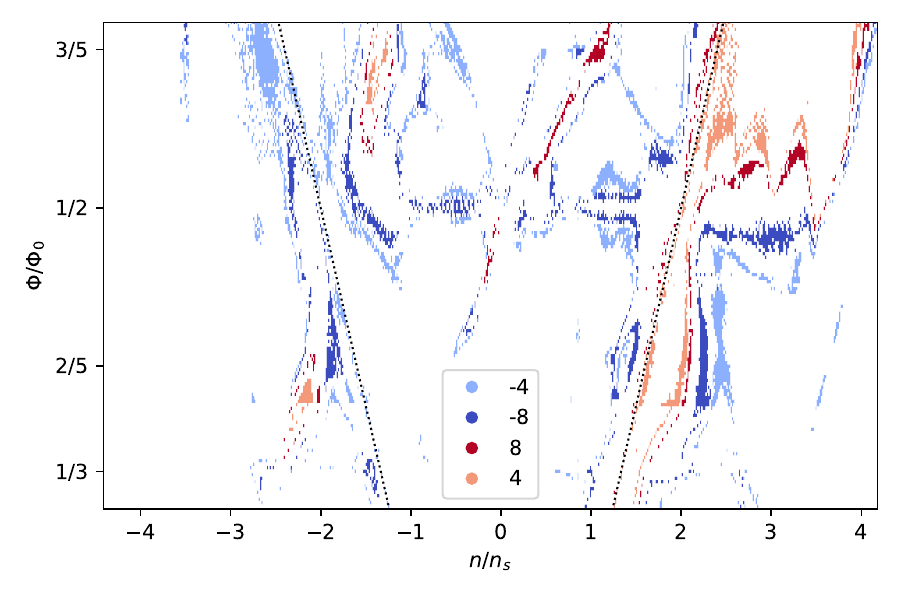}
\caption{\textbf{Integer quantum Hall plateaus.} $\pm200$ $\Omega$ windows for several different integer plateaus for contact pair 7~-~17. Dotted lines are (0, pm 4). Though the integer states are reasonably well quantized, they subtend far less of the Landau fan at these fluxes than 8/3.}
\label{fig:integers}
\end{figure*}




\begin{figure*}
\centering
\includegraphics[width=6in]{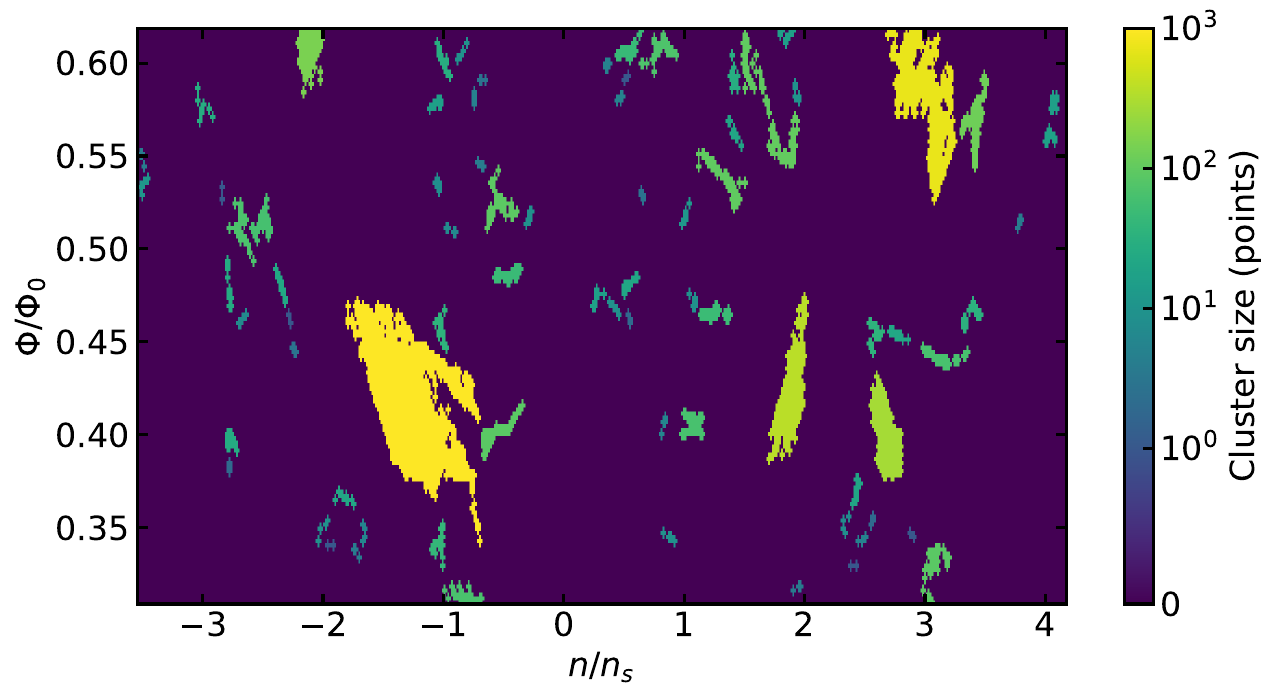}
\caption{\textbf{Clustering mask for $\pm8/3$.} 
Here we color code clusters by the number of points included in the cluster. Nearest neighbors and next-nearest neighbors that both fall within the defined window will be included in a cluster.
}
\label{fig:cluster}
\end{figure*}

\begin{figure*}
\centering
\includegraphics[width=6in]{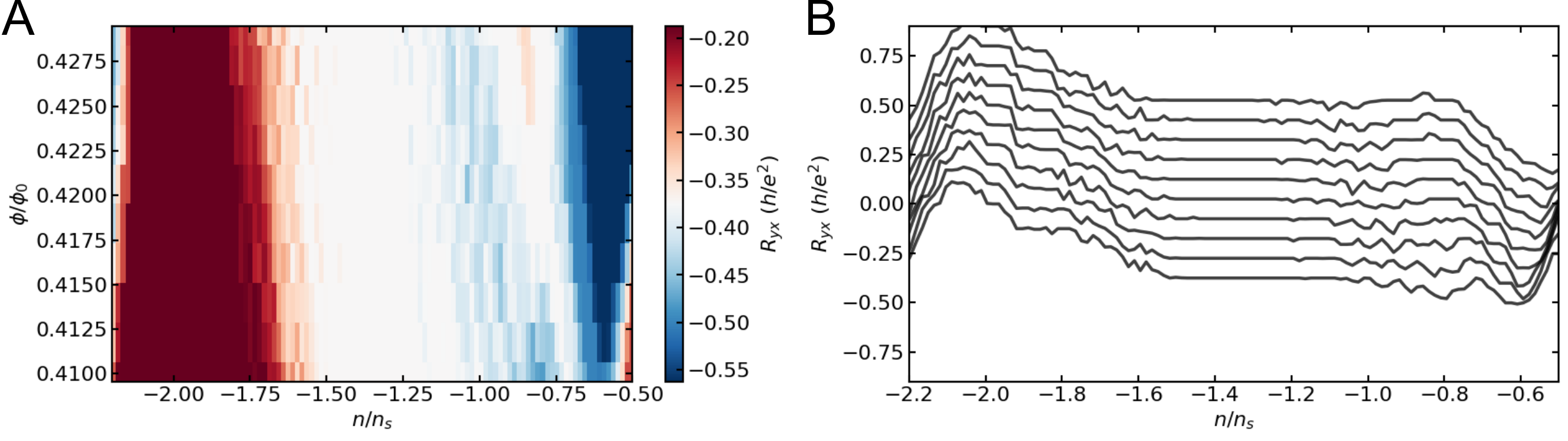}
\caption{\textbf{Possible reentrant FCI.} 
(\textbf{A}) Zoomed in colorplot of the Hall resistance in on the largest plateau of -8/3 quantization. Colorbar is centered about $3h/8e^2$, such that white regions are well quantized. (\textbf{B}) Waterfall plot of (A) with curves offset for clarity. There is a loss of quantization roughly around -1 $n/n_s$ with quantization recovered on either side. 
}
\label{fig:reentrant}
\end{figure*}

\begin{figure*}[th]
\centering
\includegraphics[width=18cm]{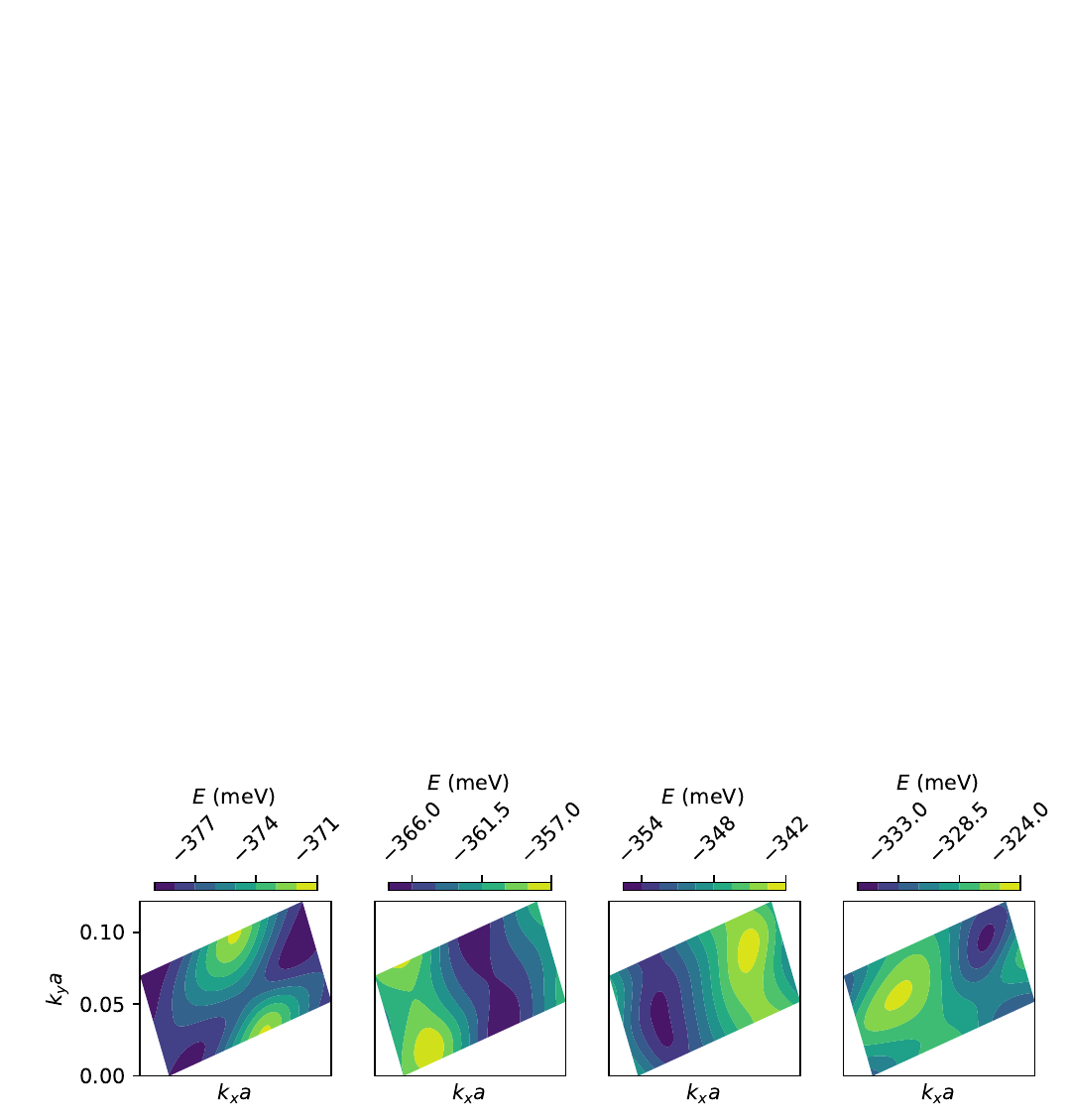}
\caption{\textbf{Computed band structure at 1/2 flux.} Computed energy levels for the four bands at half flux. The middlemost bands have regions of open orbits.}
\label{fig:halffluxbands}
\end{figure*}

\begin{figure*}
\centering
\includegraphics[width=6in]{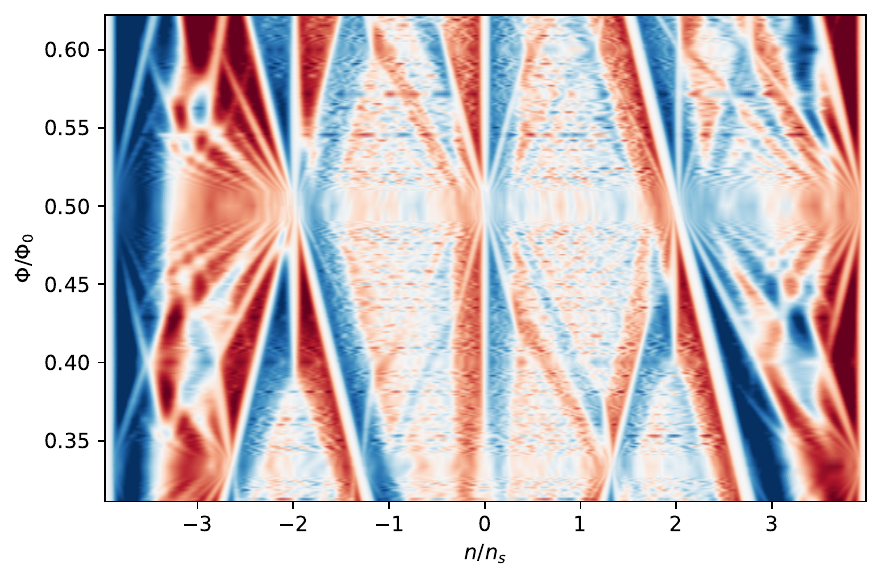}
\caption{\textbf{Spin states in computation.} Density of states for spin up minus spin down. Blue means more spin-up than spin down, and red is the opposite.}
\label{fig:densdiff}
\end{figure*}

\begin{figure*}
\centering
\includegraphics[width=6in]{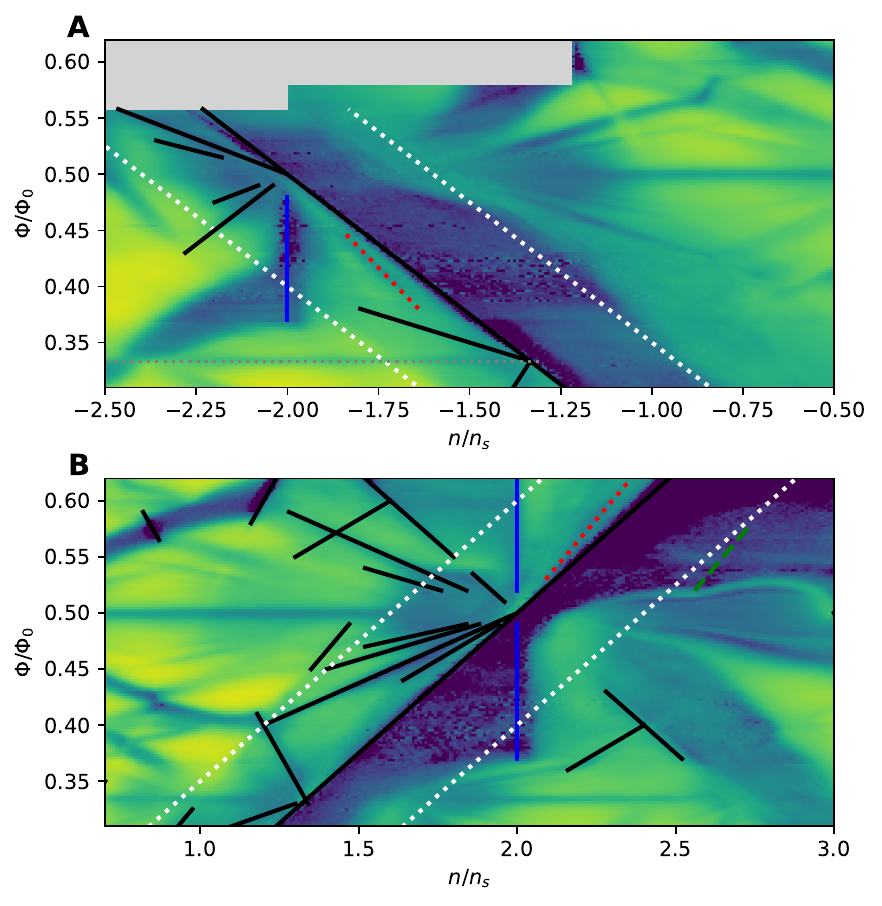}
\caption{\textbf{Spin-dependent transport detail.} Zoomed in from Fig.~\ref{fig:megafan}. On the left side of $(0, -4)$ (\textbf{A}) and $(0, 4)$ (\textbf{B}), the resistance is higher than on the right side and there are St\u{r}eda lines (black lines). Dotted white lines approximate where we start to have two spin species. See Fig.~\ref{fig:densdiff}.}
\label{fig:spindepzoom}
\end{figure*}

\end{document}